\documentclass[journal]{./IEEEtran}
\usepackage{cite}
\usepackage{soul}

\ifCLASSINFOpdf
  \usepackage[pdftex]{graphicx}
\else
\begin{center}

\end{center}
  \usepackage[dvips]{graphicx}
\fi
\usepackage{amsmath}
\usepackage{algorithm}
\usepackage{algorithmic}
\usepackage{amsfonts,amssymb,amsthm}
\usepackage{upgreek}
\usepackage[tight]{subfigure}
\usepackage{multirow}
\usepackage{tabularx}
\usepackage{color, soul}

\renewcommand{\hl}[1]{}

\begin{document}
\title{Digital Metasurface Based on Graphene: An Application to Beam Steering in Terahertz Plasmonic Antennas}
\author{Seyed Ehsan Hosseininejad, Kasra Rouhi, Mohammad Neshat, Albert Cabellos-Aparicio, Sergi Abadal and Eduard Alarc\'{o}n 
 % <-this % stops a space
%\thanks{Manuscript received XXXXXXX; revised August XXXXXX.}%
\thanks{S. E. Hosseininejad and M. Neshat are with the School of Electrical and Computer Engineering, University of Tehran, Tehran, Iran, (email: sehosseininejad@ut.ac.ir; mneshat@ut.ac.ir)}% <-this % stops a space
\thanks{K. Rouhi is with the School of Electrical Engineering, Iran University of Science and Technology, Tehran, Iran, (email: kasrarouhi@elec.iust.ac.ir)}
\thanks{E. Alarc\'{o}n, A. Cabellos-Aparicio, and S. Abadal are with the NaNoNetworking Center in Catalonia (N3Cat), Universitat Polit\`{e}cnica de Catalunya, 08034 Barcelona, Spain (e-mail: eduard.alarcon@upc.edu, acabello@ac.upc.edu, abadal@ac.upc.edu)}%
}% <-this % stops a space

\maketitle

\begin{abstract}
Metasurfaces, the two-dimensional counterpart of metamaterials, have caught great attention thanks to their powerful capabilities on manipulation of electromagnetic waves. Recent times have seen the emergence of a variety of metasurfaces exhibiting not only countless functionalities, but also a reconfigurable response. Additionally, digital or coding metasurfaces have revolutionized the field by describing the device as a matrix of discrete building block states, thus drawing clear parallelisms with information theory and opening new ways to model, compose, and (re)program advanced metasurfaces. This paper joins the reconfigurable and digital approaches, and presents a metasurface that leverages the tunability of graphene to perform beam steering at terahertz frequencies. A comprehensive design methodology is presented encompassing technological, unit cell design, digital metamaterial synthesis, and programmability aspects. By setting up and dynamically adjusting a phase gradient along the metasurface plane, the resulting device achieves beam steering at all practical directions. The proposed design is studied through analytical models and validated numerically,  showing beam widths and steering errors well below 10\textsuperscript{o} and 5\% in most cases. Finally, design guidelines are extracted through a scalability analysis involving the metasurface size and number of unit cell states. 

%Reconfigurable metasurfaces have emerged to address the well-known non-adaptivity and non-reusability constraints of conventional metasurfaces.

%Digital metasurface, a planar metamaterial can be digitally controlled, have recently enabled the realization of novel structures with unprecedented engineered functionalities. This paper proposes a plasmonic graphene-based patch as a unit cell which realizes a coding metasurface in the THz frequencies just by changing chemical potential of graphene. Based on this particle, we proposes a metasurface antenna  to digitally achieve beam steering. Results show ...     
\end{abstract}


% \fi
\IEEEpeerreviewmaketitle

\section{Introduction}
\label{sec:intro}
% The very first letter is a 2 line 
%  drop letter followed

%% 1) Definition of metasurface -- uses and applications.
\IEEEPARstart{M}{etasurfaces}, defined as artificial structures with subwavelength thickness, have enabled the realization of novel compact devices with unprecedented electromagnetic control. Frequency selectivity, absorption, anomalous reflection/transmission, polarization conversion, and focusing are among the many electromagnetic functionalities that can be achieved through the careful design of the metasurfaces \cite{Chen2016, Glybovski2016}. With such unprecedented control of the response to the impinging wave, metasurfaces have led to important breakthroughs in electromagnetic cloaking, imaging, as well as in the creation of ultra-efficient, miniaturized antennas for sensors and implantable communication devices \cite{Chen2011, ChenXZ2012, Li2017b, Vellucci2017, Tasolamprou2017, Tsilipakos2018a}.

%% 1b) In the terahertz band, efforts and applications.
A metasurface is generally defined as a planar array of periodic or quasi-periodic subwavelength elements, whose structure and coupling determine the electromagnetic function. As long as the elements remain subwavelength in size, the working principle of metasurfaces can be applied from microwaves to the visible range \cite{Glybovski2016}. Between these two extremes %(at tens of microns per unit cell) 
lies the terahertz (THz) band, for which designs have been reported to manipulate the phase, amplitude or polarization of the waves reflected or transmitted by the metasurface \cite{Qu2015, Zhang2016a, Liu2016a, Qu2017}. 

%% 2) Towards programmable metasurfaces.
The main issue of conventional metasurfaces are the lack of adaptivity and reconfigurability as, in most designs, the electromagnetic function and its scope are fixed once the unit cell is designed. In order to avoid re-designing and re-fabricating metasurfaces each time a change in frequency or functionality is required, one can introduce tunable or switchable elements in the design of unit cells \cite{Oliveri2015}. The resulting reconfigurable metasurfaces can be globally or locally tunable depending on the specific design, and better yet through appropriate control means, they can become programmable \cite{Liu2018ISCAS, Liu2018}. 

Coding metamaterials, sometimes also referred to as \emph{digital} metamaterials, are a particular type of programmable metamaterials that discretize the number of states of a unit cell \cite{liu2017concepts, Liang2015a, DellaGiovampaola2014, Cui2014}. Each state is represented by a number of bits that are used to make the actual metasurface. A desired global response is achieved through a medium profile that is not necessarily periodical. Such structure, when built using locally switchable elements, can be elegantly described as a bit or state matrix and digitally controlled through reconfigurable devices such as Field-Programmable Gate Arrays (FPGAs) \cite{Cui2014}. Several examples implementing polarization control, focusing control, or beam manipulation in the GHz range can be found in the literature \cite{Cui2014, Yang2016, Huang2017}.
% Local tunability ==> discretization through switchability ==> global coding

%% Furthermore, the concept of coding metasurfaces, realizing a multitude of functionalities by binary coding elements with controlled sequences, has been recently proposed to provide a simplified design manner of metasurfaces [15:Cui]."   

%% 3) Graphene as key element for THz devices --- metasurfaces as well. 
Graphene, with its outstanding optoelectrical properties, has been recently introduced as a key enabler of a myriad of applications in countless domains \cite{Novoselov2012, Wu2012, Low2014, Hosseininejad2017}. It is well known that graphene naturally supports Surface Plasmon Polaritons (SPP) in the terahertz band, and therefore, becomes an excellent option for the implementation of terahertz sources \cite{Jornet2014TRANSCEIVER} and antennas \cite{Correas2017}, among others. The plasmonic nature of graphene at terahertz frequencies leads to miniaturized devices \cite{LlatserComparison}, whereas its inherent tunability has been leveraged in frequency-agile or reconfigurable concepts \cite{tamagnone2012reconfigurable,  Hosseininejad2016, Hosseininejad2018EuCAP}. Some of such designs are array-based, and similar to programmable metasurfaces, they achieve reconfigurability by switching the state of its elements, i.e. \emph{tuning them in or out} \cite{Huang2012ARRAY, Xu2014MIMO, Hosseininejad2018WCNC}.

\begin{figure*}[!ht] 
\centering
{\includegraphics[width=1\textwidth]{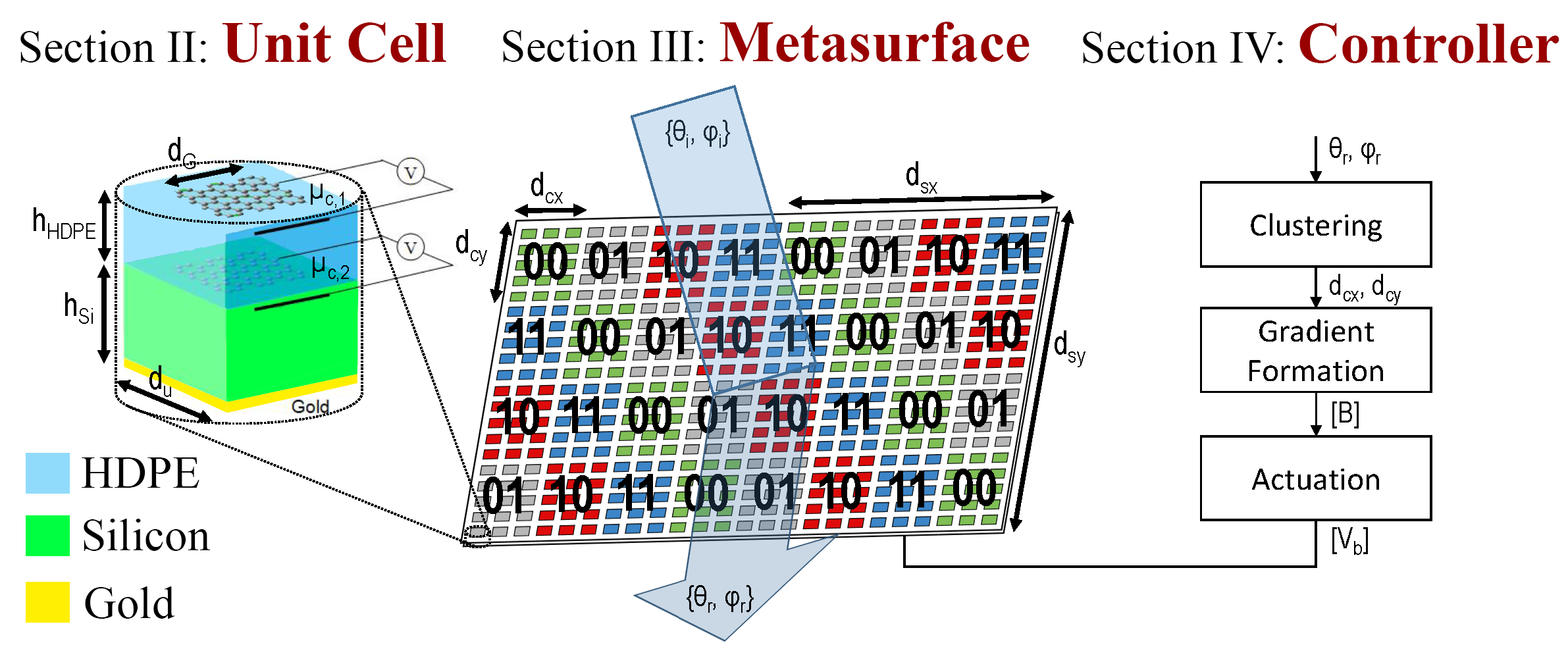}}%,natwidth=426,natheight=338
\vspace{-0.3cm}
\caption{Sketch of the programmable graphene-based digital metasurface for THz beam steering and its design flow: from the unit cell to the global controller.\label{fig:summary}}
\vspace{-0.3cm}
\end{figure*} % You must have at least 2 lines in the paragraph with the drop letter

%% 4) Graphene reconfigurable metasurfaces.
The above-mentioned properties turn graphene into a unique material for the implementation of terahertz reconfigurable metasurfaces. First explorations in this regard considered graphene reflectarrays and studied their amplitude-phase responses when tuning the chemical potential of the graphene \cite{Carrasco2013a}. By means of local tuning of the graphene elements through electrostatic biasing, the scattering profile of the reflectarray can be modified to achieve beam steering \cite{Orazbayev2017}, focusing \cite{Hosseininejad2019}, diffusive scattering \cite{Rouhi2017a}, cloaking \cite{Biswas2018} or wave vorticity control \cite{Chang2016}. \hl{These functionalities have been achieved in the microwave regime thanks to the use of} phase change materials (PCMs) \cite{Chen2015SR}, semiconductor diodes \cite{Perruisseau2010}, \hl{or microelectromechanical systems (MEMS)} \cite{Ma2014light}. \hl{However, as we approach to terahertz frequencies, diodes and MEMS become lossy and too large to be integrated within individual unit cells. On the other hand, PCMs offer limited reconfigurability as they generally switch among two states only}. \hl{With graphene}, the design can be greatly simplified and the device can be reconfigured much faster.

%Graphene terahertz metasurfaces have been recently proposed as well given its inherent tunability and good terahertz properties. 

%% 5) Contribution. Related work. Uniqueness (programmable beam steering: fast, programmable, no need for expensive and super-accurate phase shifters).
%% FOUR PILLARS: reflectarray. Digital metasurface. Graphene plasmonics. Terahertz
Although the natural switchability of graphene in the terahertz band matches perfectly with the coding metamaterial paradigm, which has been explored in \cite{Wang2015} for the first time, the lack of literature to present a clear methodology for the unit cell design based on graphene and the coding of the metasurface seems evident. To bridge this gap, this paper presents a comprehensive methodology for the design of programmable metasurfaces from the unit cell to the metasurface controller (Figure \ref{fig:summary}). The proposed methodology is then applied to develop a metasurface for fine-grained beam steering at terahertz frequencies. The metasurface acts as a reflectarray that forms dynamically reconfigurable phase gradients in the X and Y directions, through which the reflected beam can be driven to any desired direction. The unit cells of the reflectarray are based on a graphene-insulator-graphene stack that achieves wide phase tuning via electrostatic biasing of the graphene patches. With two bits per unit cell and the appropriate controller, the proposed metasurface achieves a very wide steering range with low beam width.   

The proposed metasurface is particularly suitable for wireless communication applications. In this context, the use of the lower part of the THz spectrum (our design operates at $f=2$ THz) becomes extremely attractive due to the abundance of bandwidth that allows to satisfy the extreme data rate demands of 5G networks and beyond \cite{Akyildiz2014a}. Communication in the THz band, however, requires overcoming high path losses mainly through directive antennas with very narrow beams and through the use of smart programmable reflectors \cite{Akyildiz2018, Akyildiz2016, Tan2018, Liaskos2018a, Hosseininejad2018WCNC}. It is thus fundamental that these devices be capable of steering the THz beam with high precision to track the users and avoid interrupting communication.

%The operating principle of the metasurface is the creation and dynamic reconfiguration of phase gradients in the X and Y directions, through which an arbitrary  
%The metasurface can be seen as a reflectarray that creates and dynamically reconfigures p
% We also motivate the design of the control layer required to achieve such accuracy by dimensioning     

In summary, the main contributions of this paper are:
\begin{itemize}
\item The development of a comprehensive methodology for the design of graphene-based programmable terahertz metasurfaces for beam steering, from the unit cell up to the global controller.
\item The use of the proposed methodology to design and evaluate a 2-bit coding metasurface for beam steering. Wide steering range with a sharp reflected beam and low overheads are demonstrated. The chosen frequency of operation is $f=2$ THz, within the range expected for THz wireless communication applications.
\item A scalability analysis illustrating the relation between the different design parameters and performance metrics, and uncovering several co-design opportunities.
\end{itemize}

% 6) Remainder of the paper.
The rest of this paper is organized as shown schematically in Fig. \ref{fig:summary}. Section \ref{sec:geometry} presents a design space exploration of graphene-based unit cells from the perspectives of size, chemical potential, and number of states. Section \ref{sec:coding} formulates a design flow for beam steering coding metasurfaces, which is then tested by showing the effective steering of the antenna beam in several directions. Section \ref{sec:antenna} discusses and evaluates the implementation of the scheme that actually controls and (re)programs of the metasurface. Finally, Section \ref{sec:disc} outlines the main scalability trends and co-design opportunities of the proposed design. Section \ref{sec:conclusion} concludes the paper.

\begin{figure*}[!ht] 
\centering
\subfigure[Full layer unit cell (\textsc{1L}).\label{fig:1G}]{\includegraphics[width=1\columnwidth]{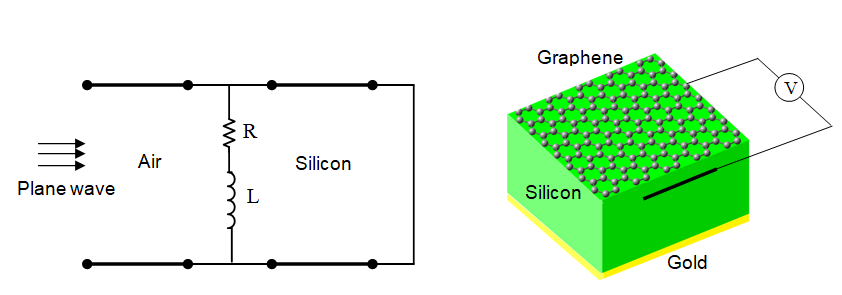}} % ,natwidth=426,natheight=338
\subfigure[Single patch unit cell (\textsc{1G}).\label{fig:1G-patch}]{\includegraphics[width=1\columnwidth]{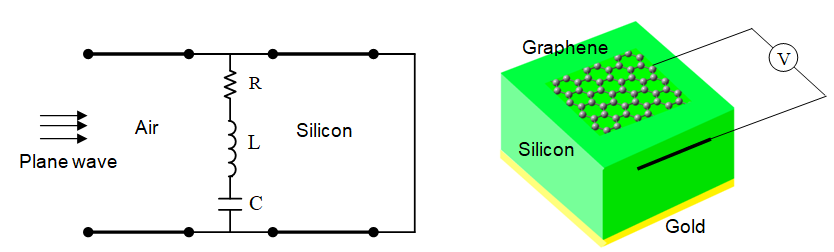}} % ,natwidth=426,natheight=484
\subfigure[Dual patch unit cell (\textsc{2G}).\label{fig:2G-patch}]{\includegraphics[width=1\columnwidth]{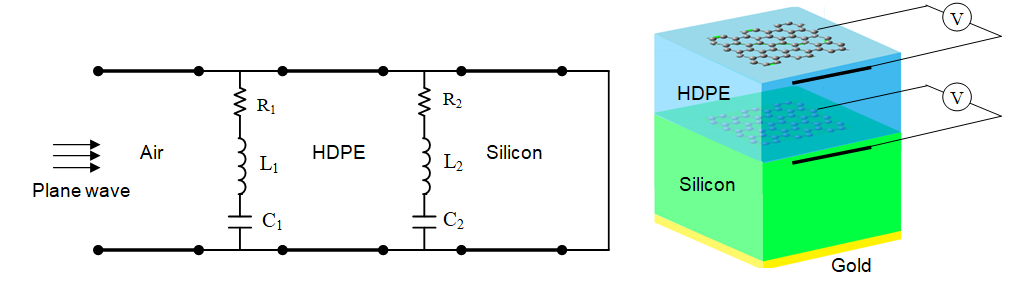}}  % ,natwidth=426,natheight=484
\vspace{-0.2cm}
\caption{A schematic representation of the graphene unit cells with their respective equivalent circuit models.}
\label{fig:unit cells}
\vspace{-0.3cm}
\end{figure*} % You must have at least 2 lines in the paragraph with the drop letter

\section{Graphene-based unit cell}
\label{sec:geometry}
The design of any metasurface starts with its most basic building block, namely, the unit cell. For beam manipulation, we need to provide a unit cell with the ability of controlling the phase response over a wide range of values \cite{Qu2017}. Moreover, since the proposed device acts as a reflectarray, the unit cell needs to yield a high reflection amplitude at all times. To enable dynamic reconfigurability, it is necessary that both objectives can be met without physically changing any geometry. In this paper, reconfigurability is achieved at THz frequencies by means of the electrostatic tuning of graphene. % FIX: advantage of making this with graphene over the others? Plasmonic, better for steering.

%The aim of this section is to deliver a design that achieves both objectives by electrically tuning graphene. This method flexible and easier to scale than by physically changing the size of the unit cell \cite{}. %Finding such a scheme is a challenging task that is addressed in this section. 

%with high reflection amplitude is a challenging task that is addressed here.
%Looking at a reflective metasurface as a beam steering antenna to give a radiation pattern in a specific direction, the reflection amplitude should be close to one and its phase varying with a linear gradient. Consequently, finding a unique particle with the ability of controlling the phase with high reflection amplitude is a challenging task that is addressed here.

\subsection{Graphene Modeling}
We analyze different unit cells that leverage the tunability of graphene to achieve the desired phase variation with reasonable losses and without the need of changing any geometry. To drive the design and to perform an accurate evaluation of different proposals, we model graphene as an infinitesimally thin sheet with surface impedance $Z = 1/\sigma(\omega)$, where $\sigma(\omega)$ is the frequency-dependent conductivity of graphene. The complex conductivity is given by
\begin{equation}
\sigma\left(\omega\right)=\frac{2e^{2}}{\pi\hbar}\frac{k_{B}T}{\hbar}\ln\left[2\cosh\left[\frac{\mu_{c}}{2k_{B}T}\right]\right]\frac{i}{\omega+i\tau^{-1}},\label{eq:sigma_graphene}
\end{equation}
where $e$, $\hbar$ and $k_{B}$ are constants corresponding to the charge of an electron, the reduced Planck constant and the Boltzmann constant, respectively \cite{Hanson2008}. Variables $T$, $\tau$ and $\mu_{c}$ correspond to the temperature, the relaxation time and the chemical potential of the graphene layer. Note that this expression neglects the edge effects of the graphene and considers that the Drude-like intraband contribution dominates, which are experimentally validated assumptions at the sizes and frequencies considered in this work \cite{AbadalTCOM}.

On the one hand, the phase control in graphene metasurface is achieved via changes in its complex conductivity when biased -- the effect that can be modeled through the chemical potential value $\mu_{c}$. The chemical potential can be controlled through electrostatic biasing, and therefore, we can meet the phase change requirement. On the other hand, the amplitude response depends on the losses within graphene, which are mostly influenced by the relaxation time value $\tau$. Note that the relaxation time is proportional to the carrier mobility, which depends on the quality of the material. For the purpose of this work, losses will be affordable as long as the carrier mobility of the graphene sheets is on the order of 10,000 cm\textsuperscript{2}V\textsuperscript{-1}s\textsuperscript{-1}, which is achievable with current fabrication and encapsulation techniques \cite{Banszerus2015}. Thus, the amplitude requirement can be met as well.   
%As a matter of fact, graphene provides controlling the phase of the reflected wave at each unitcell using changing its complex conductivity when biased by electrostatic field.  

\subsection{Unit Cell Design}
Figure \ref{fig:unit cells} shows a schematic representation of the proposed unit cells together with their approximate equivalent circuit models. We numerically simulate different designs in CST Microwave Studio \cite{CST} to obtain the amplitude and phase responses. Then, through the equivalent circuit models, we verify the results of the numerical approach, and reason about the behavior of different unit cells. In all cases, we assume a lateral size of $d_{u} = 20\,\upmu$m. This value is around $\lambda_{0}/8$ for the targeted frequency of operation (2 THz), enough to provide the subwavelength behavior required in the metasurface. The relaxation time of graphene is assumed to be $\tau = 0.6$ ps, which is compatible with the carrier mobility requirements mentioned above. 
% Smaller would be unneccessary complexity, and too big would lose the principle of metasurfaces. 

\begin{figure*}[!ht] 
\centering
\label{fig:1G1}{\includegraphics[width=2\columnwidth]{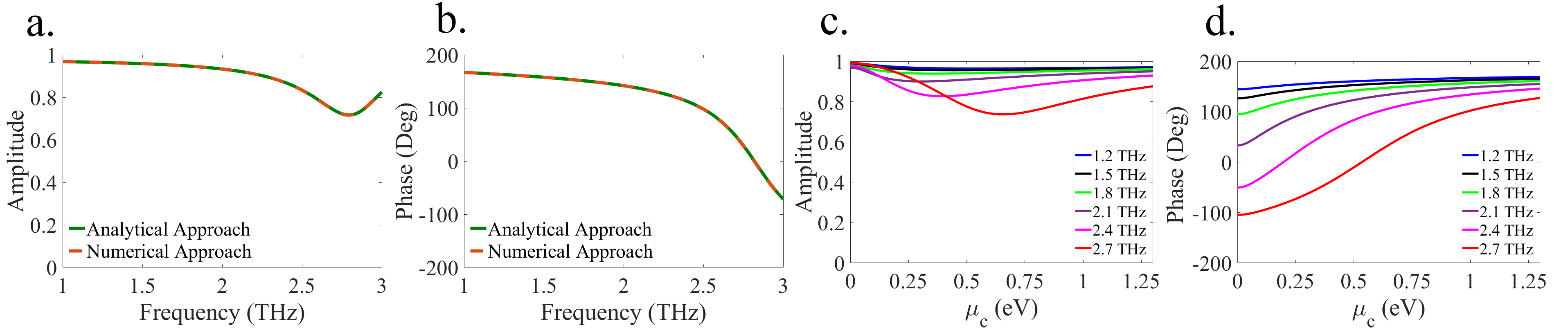}} %,natwidth=426,natheight=338
\vspace{-0.2cm}
\caption{a) Amplitude and b) phase responses of reflection coefficient for the \textsc{1L} unit cell. Effect of chemical potential variation in c) amplitude and d) phase for various frequency. Unless noted, $E_{F} = 0.7$ eV, $\tau = 0.6$ ps, and $f = 2$ THz. \label{fig:results1G}}
\vspace{-0.1cm}
\end{figure*}

The first unit cell (Fig. \ref{fig:1G}) consists of a fully covered layer of graphene on top of a silicon substrate with refractive index $n_{Si} = 3.45$ and thickness $d_{Si} = 10 \upmu$m along with a metallic ground plane on the backside. In such a unit cell, graphene is a lossy medium that can be modeled through an $RL$ circuit. Fig. \ref{fig:results1G} plots the amplitude and phase responses of reflection coefficient for such unit cell versus frequency and the chemical potential. It is observed that the amplitude response is within an acceptable range, whereas the phase range is not wide enough --around 135\textsuperscript{o} at 2 THz assuming a maximum chemical potential range of 1 eV. A very good agreement is also obtained between the numerical results and the equivalent circuit model.  
%First unit cell: changes of around 180 degrees, not enough for beam steering.

\begin{figure*}[!ht] 
\centering
\label{fig:1Gpatch1}{\includegraphics[width=2\columnwidth]{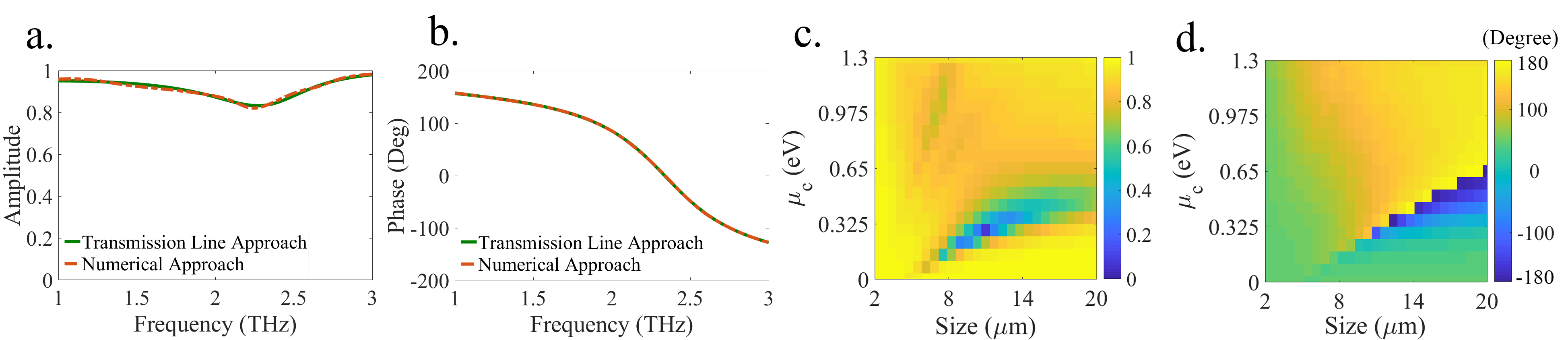}} %,natwidth=426,natheight=338
\vspace{-0.2cm} 
\caption{a) Amplitude and b) phase responses of reflection coefficient for the \textsc{1G} unit cell. Unless noted, $E_{F} = 0.2$ eV, $\tau = 0.6$ ps, $d_{G} = 16 \upmu$m, and $f = 2$ THz. Effect of chemical potential and patch size variation in c) amplitude and d) phase for a constant frequency $f = 2$ THz.\label{fig:results1Gp}}
\vspace{-0.1cm}
\end{figure*} 
% FIX: Ehsan - Interpolation in heat maps.

\begin{figure*}[!ht] 
\centering
\label{fig:2Gpatch1}{\includegraphics[width=2\columnwidth]{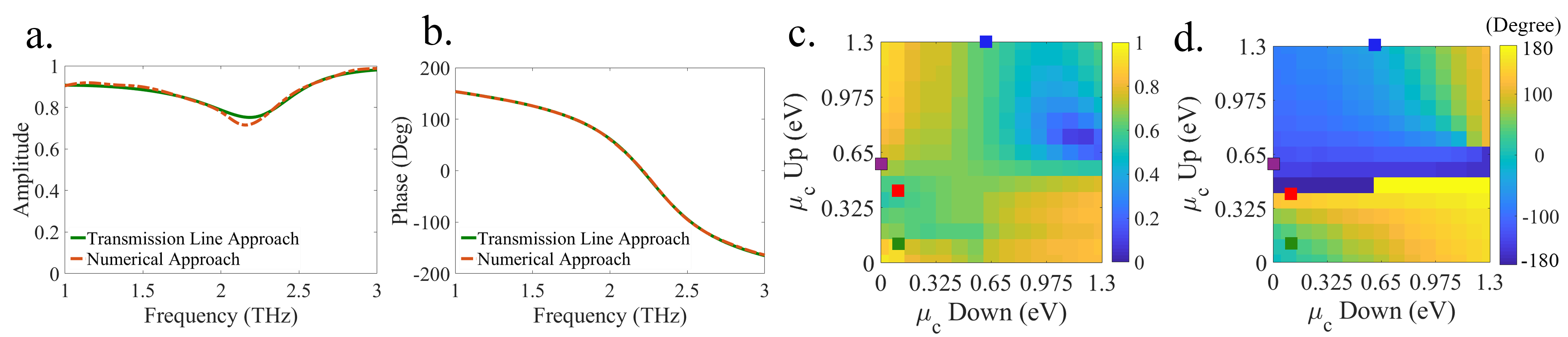}} %,natwidth=426,natheight=338
\vspace{-0.2cm}
\caption{a) Amplitude and b) phase responses of reflection coefficient for the \textsc{2G} unit cell. In the frequency response figures, $E_{F} = 0.2$ eV, $\tau = 0.6$ ps, $d_{G} = 16 \upmu$m. Effect of top and bottom layers chemical potential in c) amplitude and d) phase for a constant frequency (the four chosen coding states used in the simulation are marked by different signs). In these figures, $\tau = 0.6$ ps, $d_{G} = 12 \upmu$m and $f = 2$ THz. \label{fig:results2Gp}}
\vspace{-0.2cm}
\end{figure*} 
% FIX: Ehsan - Interpolation in heat maps.

The second unit cell (Fig. \ref{fig:1G-patch}) consists of a graphene patch that partially covers the unit cell. The substrate and ground plane remain unchanged. In this case, a capacitance is introduced to model the coupling effects generated between the edges of adjacent graphene patches. As shown in Fig. \ref{fig:results1Gp}, the size of the graphene patch provides an extra degree of freedom to deliver the target amplitude and phase responses. Exploring the \{chemical potential, patch size\} design space, we observe that there is a tradeoff between the amplitude and phase variation. For a patch size of 6 $\upmu$m, the phase range covers almost 360\textsuperscript{o} with less than 0.8 eV chemical potential variation. However, the amplitude response also has a very large variation, which discourages the choice of such design point. A patch size of 8 $\upmu$m or larger provides a better amplitude response with a reasonable phase change variation. 
%Second unit cell, with capacitance, achieves wider phase change. 
%From figure of unit cell 2: high reflection coefficient for all chemical potentials and enough phase change.

The third unit cell (Fig. \ref{fig:2G-patch}) is composed of a graphene-insulator-graphene stack placed over the substrate along with a ground plane on the backside. High-density polyethylene (HDPE) is chosen as the insulator due to its particularly low losses in the terahertz band \cite{Zhou2014a}. The refractive index of the insulator is $n_{HDPE} = 1.54$ and its thickness is $d_{HDPE} = 4 \upmu$m. The equivalent circuit model of this structure consists of two parallel $RLC$ cells representing each of the graphene sheets. As shown in Fig. \ref{fig:results2Gp}, this unit cell achieves a much wider phase variation, and by addressing each graphene patch independently, provides an extra degree of freedom to choose the states of the metasurface.

\hl{According to the formulation presented in [], the initial values for RLC model parameters are estimated. For the dual patch unit cell, due to the coupling effect between two graphene layers, small changes can occur in the RLC parameters which are optimized by genetic algoritm. The parameters for three unit cell are calculated respectively as....}

Regarding fabrication feasibility of the proposed low-profile structure (10 $\upmu$m substrate), there are advanced Silicon substrate thinning techniques that can be used to achieve an ultra-thinning down to 4 $\upmu$m without damage occurred due to thinning processes \cite{kim2014ultra}. 

\subsection{Unit Cell Discrete States}
\label{sec:states}
The results above show the response of the metasurface for a continuous range of chemical potentials. However, in order to design a bit-programmable metasurface, we need to discretize the potentials to obtain a finite set of addressable states. 

The first decision concerns the number of target states, and thus, the number of bits required to address a unit cell. Here, the number of states will determine the phase difference between consecutive unit cell states. For the application at hand, there is a relation between the phase difference and the steering resolution, i.e., the angle difference between consecutive achievable beam directions (see Sec. \ref{sec:disc} for details). Therefore, the number of bits must be chosen carefully. 

\begin{figure}[!t] 
\centering
\label{fig:1code1}{\includegraphics[width=0.7\columnwidth]{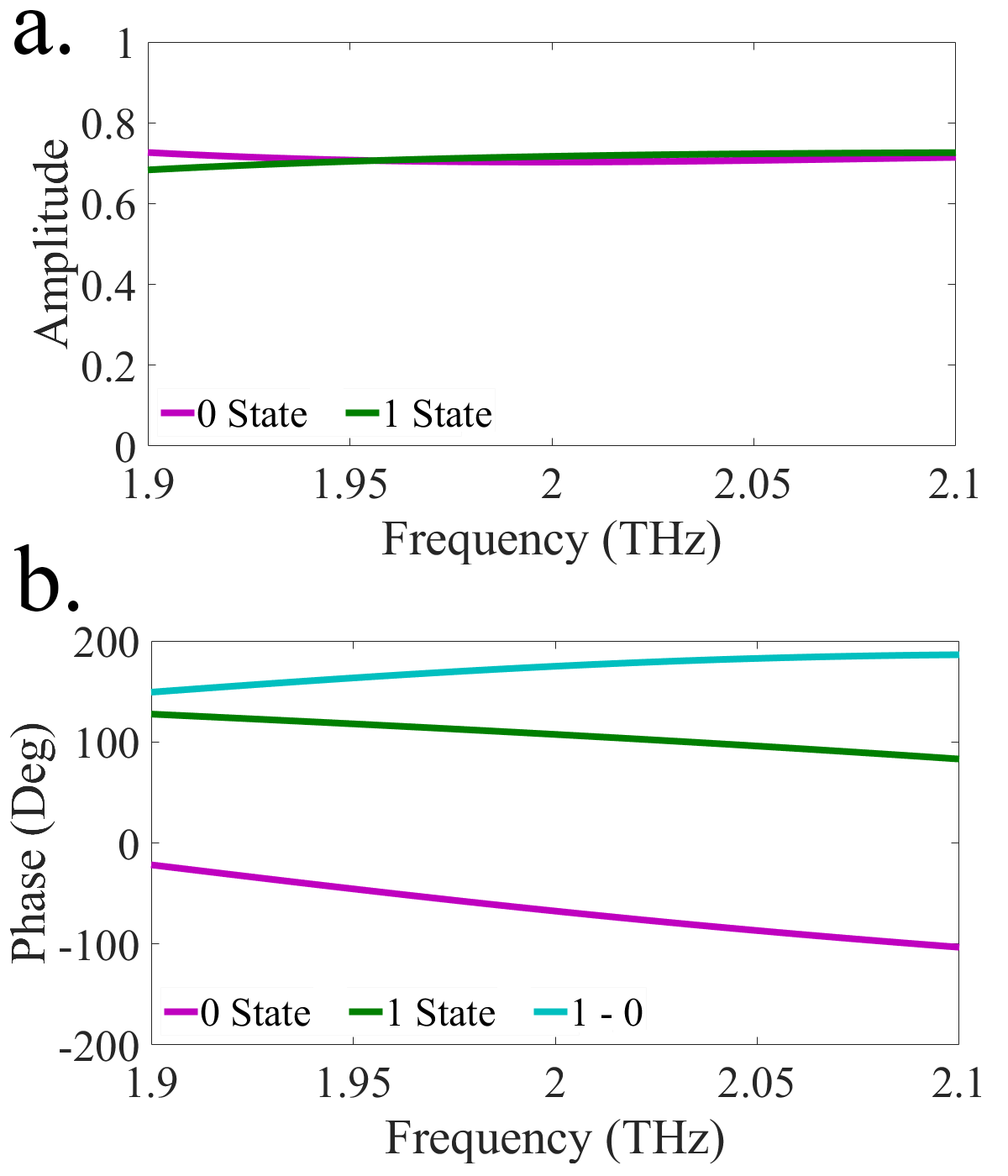}} %,natwidth=426,natheight=338
\vspace{-0.2cm}
\caption{a) Amplitude and b) phase of reflection coefficient for 1-bit digital metasurface by \textsc{1G} unit cell. \label{fig:1code}}
\vspace{-0.2cm}
\end{figure}

Let us now exemplify this process by deriving the states required for the metasurface to work at 2 THz. We start by addressing the  \textsc{1G} unit cell with a single bit. From the design space exploration shown in Fig. \ref{fig:results1Gp}, we choose design points that have high amplitude and a phase difference of approximately 180\textsuperscript{o}. A good choice is  $d_{G} = 8 \upmu$m with $\mu_{c} = \{0.6, 1.28\}$ eV corresponding to a bit combination of $B=\{0,1\}$. The resulting amplitude and phase responses, illustrated in Fig. \ref{fig:1code}, provide a constant reflection coefficient of around 0.7 and deliver the targeted 180\textsuperscript{o} phase shift. 

Two-bit coding leads to a phase shift resolution of 90\textsuperscript{o} and would improve the beam steering accuracy substantially. The  \textsc{1G} unit cell, however, barely meets the amplitude and phase shift requirements with a 90\textsuperscript{o} resolution. With $d_{G} = 8 \upmu$m, there is no combination of chemical potentials capable of avoiding the region of low amplitude around 0.9 eV. For larger patch sizes, the phase response is not wide enough to accommodate two bits. For three or more bits, this unit cell would not be suitable for beam steering, at least for the relaxation time values and geometry considered in this work.

 % You must have at least 2 lines in the paragraph with the drop letter

\begin{figure}[!t] 
\centering
\label{fig:2code1}{\includegraphics[width=0.7 \columnwidth]{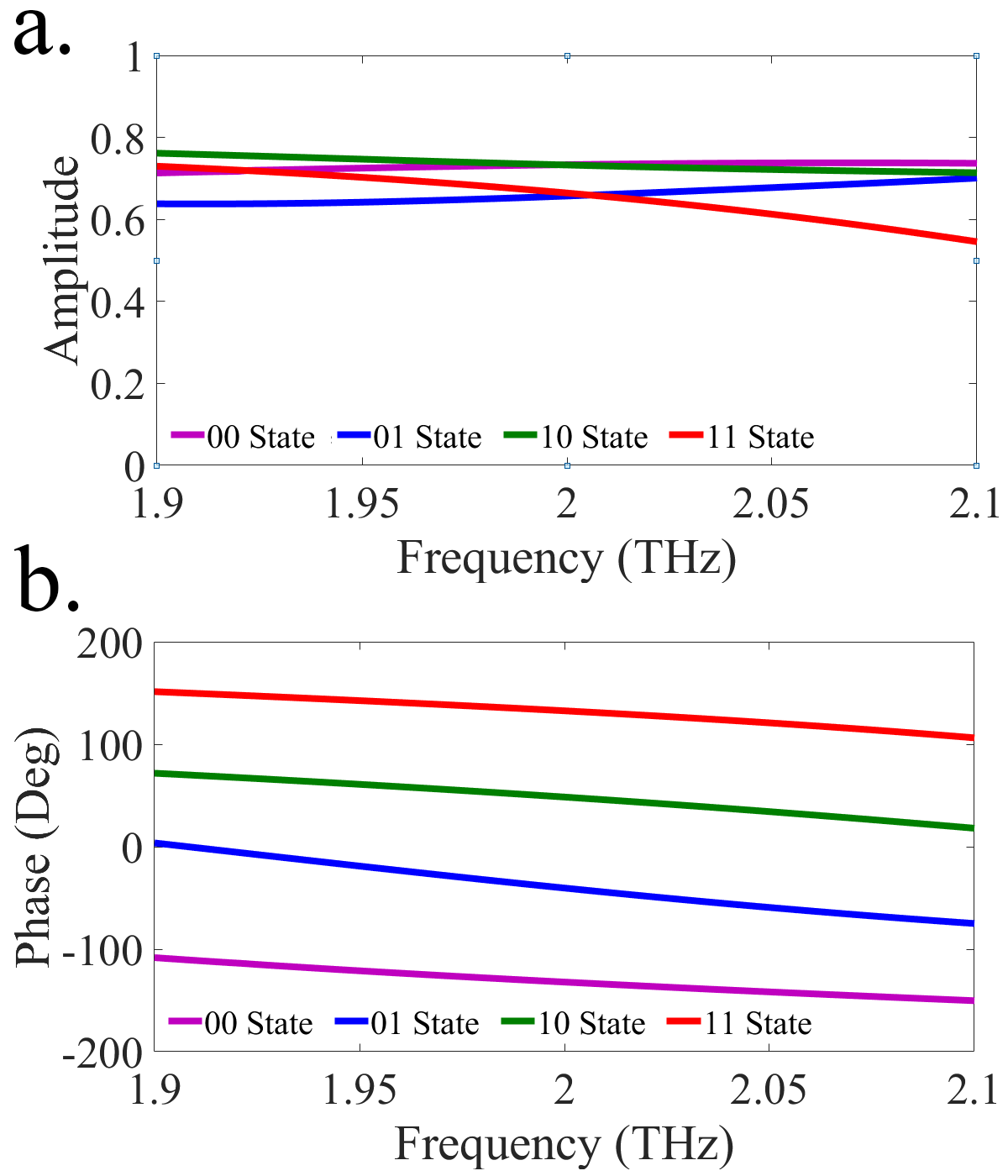}} %,natwidth=426,natheight=338
\vspace{-0.2cm}
\caption{a) Amplitude and b) phase of reflection coefficient for 2-bit digital metasurface in \textsc{2G} unit cell. \label{fig:2code}}
\vspace{-0.2cm}
\end{figure}

Alternatively, the \textsc{2G} unit cell offers much more freedom and is capable of accommodating two or more bits. Addressing the \textsc{2G} unit cell with two bits, one can find suitable design points with $d_{G}=12 \upmu$m. Using the design space exploration from Fig. \ref{fig:results2Gp}, good performance is obtained for the following up-layer and down-layer chemical potentials, respectively: $\mu_{c,1} = \{0.6, 1.3, 0.1, 0.4\}$ eV and $\mu_{c,2} = \{0, 0.6, 0.1, 0.1\}$ eV corresponding to the bit combinations $B = \{00, 01, 10, 11\}$. It is observed in Fig. \ref{fig:2code} that these states consistently achieve a reflection coefficient around 0.7 and a phase difference of 90\textsuperscript{o} covering the whole phase space. In Section \ref{sec:antenna}, we discuss how to electronically achieve these states.

%Table \ref{tab:elements} shows the extracted values of circuit elements.

%\begin{table}[!t] 
%\caption{Values of Equivalent Circuit Models.}
%\vspace{-0.2cm}
%\label{tab:elements}
%\footnotesize
%\centering
%\begin{tabular}{|c|ccc|} 
%\hline
%   & Full Layer & Single Patch & Dual Patch \\ \hline
%R1 &  \hl{100} $\Omega$ & 192.85 $\Omega$ & 20.32 $\Omega$ \\
%L1 &  \hl{100} & 63.49 pH & 0.01 pH \\
%C1 &  - & 1.36 fF &  0.1 fF \\
%R2 &  - &  -  & 35.97 $\Omega$  \\
%L2 &  - &  -  & 26.26 pH \\
%C2 &	- &  -  & 3.38 fF \\
%\hline
%\end{tabular}
%\vspace{-0.3cm}
%\end{table}

 % You must have at least 2 lines in the paragraph with the drop letter

\section{Coding Metasurface Terahertz Antenna}
\label{sec:coding}
To illustrate the design approach of a terahertz metasurface for beam steering application, a metasurface including $M\times N$ controllable unit cells is considered. Our design allows one to introduce a phase gradient by smartly changing the chemical potential $\mu_{c}$ of the graphene sheets from one unit cell to another. In this case, we need to use the generalized reflection law to evaluate the response of the metasurface \cite{yu2011light}. 

In the following, we first derive the conditions required to achieve beam steering to a desired direction $\{\theta_{r}, \phi_{r}\}$ in Section \ref{sec:formula}. Then, we define the design and configuration flow to achieve such the desired direction by our proposed design in Section \ref{sec:flow}. Finally, we evaluate the performance of the proposed metasurface in Section \ref{sec:evaluation}.

\subsection{Generalized Reflection Law Formulation}
\label{sec:formula}
\begin{figure}[!t] 
\centering
{\includegraphics[width=1\columnwidth]{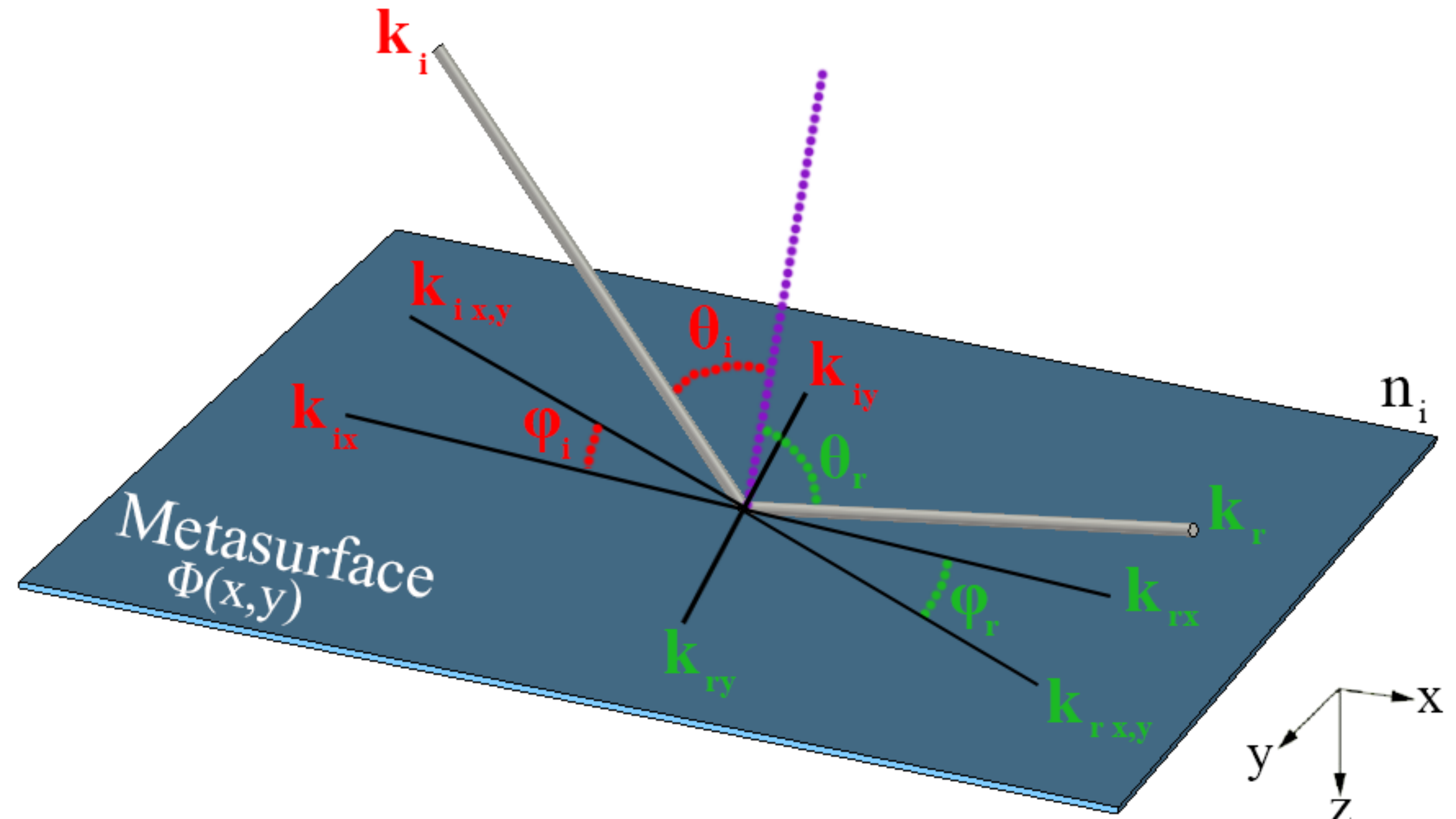}}
\vspace{-0.3cm}
\caption{Coordinate system used in the formulations of generalized reflection law.\label{fig:coor}}
\vspace{-0.1cm}
\end{figure}

Consider a reflective metasurface under illumination of an incident plane wave at elevation angle $\theta_{i}$ and azimuth angle $\phi_{i}$ according to the coordinate system shown in Fig. \ref{fig:coor}. The incident wave vector $k_{i}$ can be written as
\begin{equation}
k_{i} = k_{ix} \hat{x} + k_{iy} \hat{y} + k_{iz} \hat{z} 
\end{equation}
where $\{k_{ix}, k_{iy}, k_{iy}\}$ are the wave vector coordinates, \hl{given} by 
\begin{equation}
%\[\left\{ 
\begin{array}{l} \label{eq:ki}
k_{ix} = k_{i} \sin{\theta_{i}}\cos{\phi_{i}} = k_{0} n_{i} \sin{\theta_{i}}\cos{\phi_{i}} \\
k_{iy} = k_{i} \sin{\theta_{i}}\sin{\phi_{i}} = k_{0} n_{i} \sin{\theta_{i}}\sin{\phi_{i}} \\
k_{iz} = k_{i} \cos{\theta_{i}} = k_{0} n_{i} \cos{\theta_{i}} 
\end{array} %\right.\]
\end{equation}
The same formulation can be applied to the reflected wave vector $k_{r}$ given the elevation angle $\theta_{r}$ and azimuth angle $\phi_{r}$ of the reflected wave.

Assuming that the metasurface imposes the phase profile $\Phi(x,y)$, we assign it the virtual wave vector $k_{\Phi}$ so that 
\begin{equation}
\label{eq:kphi}
k_{\Phi} = k_{\Phi x} \hat{x} + k_{\Phi y} \hat{y} = \frac{d\Phi}{dx}\hat{x} + \frac{d\Phi}{dy}\hat{y} = \nabla \Phi_{x} \hat{x} + \nabla \Phi_{y} \hat{y} 
\end{equation}
where $\nabla_{x} \Phi = \tfrac{d\Phi}{dx}$ and $\nabla_{y} \Phi = \tfrac{d\Phi}{dy}$ are the phase gradients along the $x$ and $y$ directions, respectively. 

Applying the boundary conditions of the tangential components of the electromagnetic fields, the momentum conservation law for wave vectors can be expressed as 
\begin{equation}\label{eq:BC}
%\[\left\{ 
\begin{array}{l}
k_{ix} + k_{\Phi x} = k_{rx} \\
k_{iy} + k_{\Phi y} = k_{ry} 
\end{array} %\right.\]
\end{equation}
and substituting \eqref{eq:ki} and \eqref{eq:kphi} in \eqref{eq:BC} yields
\begin{equation}\label{eq:dphi}
%\[\left\{ 
\begin{array}{l}
k_{i} \sin{\theta_{i}}\cos{\phi_{i}} + \frac{d\Phi}{dx} = k_{r} \sin{\theta_{r}}\cos{\phi_{r}} \\
k_{i} \sin{\theta_{i}}\sin{\phi_{i}} + \frac{d\Phi}{dy} = k_{r} \sin{\theta_{r}}\sin{\phi_{r}}
\end{array} %\right.\]
\end{equation}

By mathematically simplifying the above equations as shown in the Appendix, the reflected elevation angle $\theta_{r}$ and azimuth angle $\phi_{r}$ are obtained as
\begin{equation}
\label{eq:simp}
\begin{array}{l}
\theta_{r} = \arcsin{\frac{\sqrt{(k_{i} \sin{\theta_{i}}\cos{\phi_{i}} + \frac{d\Phi}{dx})^{2} + (k_{i} \sin{\theta_{i}}\sin{\phi_{i}} + \frac{d\Phi}{dy})^{2}}}{k_{r}}} \\
\phi_{r} = \arctan{\frac{k_{i} \sin{\theta_{i}}\sin{\phi_{i}} + \frac{d\Phi}{dy}}{k_{i} \sin{\theta_{i}}\cos{\phi_{i}} + \frac{d\Phi}{dx}}}
\end{array}
\end{equation}

When the metasurface is illuminated by a normally incident wave ($\theta_{i}=\phi_{i}=0$), and assuming air as the medium of the incident and reflected wave, we can simplify the formulas as
\begin{equation}
\label{eq:reflected}
\begin{array}{l}
\theta_{r} = \arcsin{\frac{\sqrt{(\nabla_{x}\Phi)^{2}+(\nabla_{y}\Phi)^{2}}}{k_{0}}} \\
\phi_{r} = \arctan{\frac{\nabla_{y}\Phi}{\nabla_{x}\Phi}},
\end{array}
\end{equation}
which relates the phase gradient in the metasurface to the direction of the reflected wave.

\subsection{Design Flow}
\label{sec:flow}
Using \eqref{eq:reflected}, the reflected angles for all phase profiles of the metasurface can be calculated. The design of the metasurface can be then thought as an inversion process. We need to estimate the necessary phase profile to achieve the desired elevation and azimuth angles of the reflected wave. To this end, and again assuming a normal incident plane wave and air as the medium, we can rearrange \eqref{eq:dphi} as   
\begin{equation}
\label{eq:dphi2}
\begin{array}{l}
dx = \frac{\lambda_{0} d\Phi}{2\pi\cos{\phi_{r}}\sin{\theta_{r}}} \\
dy = \frac{\lambda_{0} d\Phi}{2\pi\sin{\phi_{r}}\sin{\theta_{r}}}
\end{array}
\end{equation}
where $d\Phi$ describes the phase difference between adjacent unit cell states. Starting from here, the design methodology requires the knowledge of the unit cell dimensions and the number of states. The design flow is as follows,

\vspace{0.1cm}
\noindent
\textbf{1. Obtaining the cluster size (in $\upmu$m):} Assume that the coding metasurface can choose among $2^{n}$ states for each unit cell (referred to as $n$-bit coding). In this case, the granularity of the gradient is $d\Phi = \pi/2^{n-1}$. Therefore, the lateral dimensions of the required cluster of unit cells ($d_{cx}$ and $d_{cy}$), as shown in Fig. \ref{fig:summary}, are obtained by substituting $d\Phi$ in Eq. \eqref{eq:dphi2},
\begin{equation}
\label{eq:dphi3}
\begin{array}{l}
d_{cx} = 
%dx(d\Phi=\pi/2^{n-1}) = 
\frac{\lambda_{0}}{2^{n}\cos{\phi_{r}}\sin{\theta_{r}}} \\
d_{cy} = 
%dy(d\Phi=\pi/2^{n-1}) = 
\frac{\lambda_{0}}{2^{n}\sin{\phi_{r}}\sin{\theta_{r}}}.
\end{array}
\end{equation}
It is worth noting that the results in \eqref{eq:dphi2} can be negative if $d\Phi$ becomes negative, which implies that the gradient needs to be reversed at the coding stage. We will later see that the algorithm for metasurface coding already takes the direction of the gradient into consideration. Also, the value of $\phi_{r}$ determines the difference between $d_{cx}$ and $d_{cy}$, which impacts on the shape of the reflected beam. The larger difference results in the more elliptical shape of the reflected beam.

\vspace{0.1cm}
\noindent
\textbf{2. Obtaining the cluster size (in number of unit cells):} The nature of a metasurface, consisting an array of unit cells, dictates the discretization of  space. Therefore, the values of $d_{cx}$ and $d_{cy}$ needs to be approximated in an integer number of unit cells. For this purpose, we consider that the number of unit cells in the $x$ and $y$ directions, designated by $c_{x} \in \mathbb{Z}$ and $c_{y} \in \mathbb{Z}$, respectively, are rounded as 
\begin{equation}
\label{eq:cluster}
c_{x} = \lfloor \frac{d_{cx}}{d_{u}} \rceil \,\, ,\,\, c_{x} = \lfloor \frac{d_{cy}}{d_{u}} \rceil.
\end{equation}

\begin{figure}[!t] 
\centering
{\includegraphics[width=0.8\columnwidth]{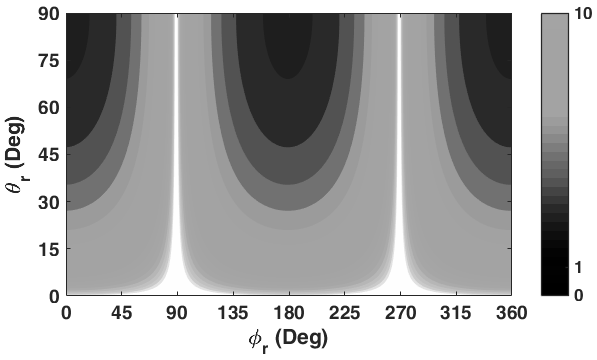}}
\vspace{-0.4cm}
\caption{Absolute value of $c_{x}$ as a function of the desired direction of reflection. 
\label{fig:cx}}
\vspace{-0.1cm}
\end{figure}

Figure \ref{fig:cx} shows the absolute value of $c_{x}$ as a function of the target direction for a normally incident plane wave. It is observed that $c_{x}$ becomes arbitrarily large as the reflected angle approaches $\theta_{r} = 0$ (white areas of the figure). This is consistent with the fact that such direction implies specular reflection, which can only be realized with a homogeneous surface, i.e. zero gradient. For directions approaching $\phi=\pi/2,~3\pi/2$, $c_{x}$ also becomes large because the gradient is only needed in the $y$ axis. On the contrary, $c_{x}$ approaches zero in the co-planar directions, where an infinite gradient would be required. The black area in Fig. \ref{fig:cx} denotes $c_{x} < 1$, which is unfeasible.

\vspace{0.1cm}
\noindent
\textbf{3. Obtaining the size of the super unit cell:} To calculate the size of the super unit cell, designated by $s_{x} \in \mathbb{Z}$ and $s_{y} \in \mathbb{Z}$, in number of unit cells, one needs to apply
\begin{equation}
\label{eq:dphi4}
s_{x} = 2^{n}c_{x} \,\, ,\,\, s_{y} = 2^{n}c_{y}.
\end{equation}

%A phase-gradient metasurfaces is able to introduce a reconfigurable-beam antenna. The concept is based on the generalized snell's law; its reflection form can be expressed by the following formulas:
%
%\[\left\{ \begin{array}{l}
%\sin ({\theta _r}) - \sin ({\theta _i}) = \frac{1}{{{n_i}{k_0}}}\frac{{d\varphi }}{{dx}}\\
%\cos ({\theta _r})\sin ({\varphi _r}) = \frac{1}{{{n_r}{k_0}}}\frac{{d\varphi }}{{dy}}
%\end{array} \right.\]

\subsection{Evaluation}
\label{sec:evaluation}
In this section, we evaluate the proposed metasurface both numerically and analytically. We numerically model the four states of the $2G$ unit cell in CST \cite{CST}, and apply the formulation developed above to assign the states to different unit cells of an $M\times N$ metasurface. Then, we obtain the response of the metasurface in the form of the far field pattern produced by a normally incident plane wave. We assume that the beam covers the whole metasurface.
% Time domain solver

\begin{figure*}[!t] 
\centering
\label{fig:Meta1N}{\includegraphics[width=1.8\columnwidth]{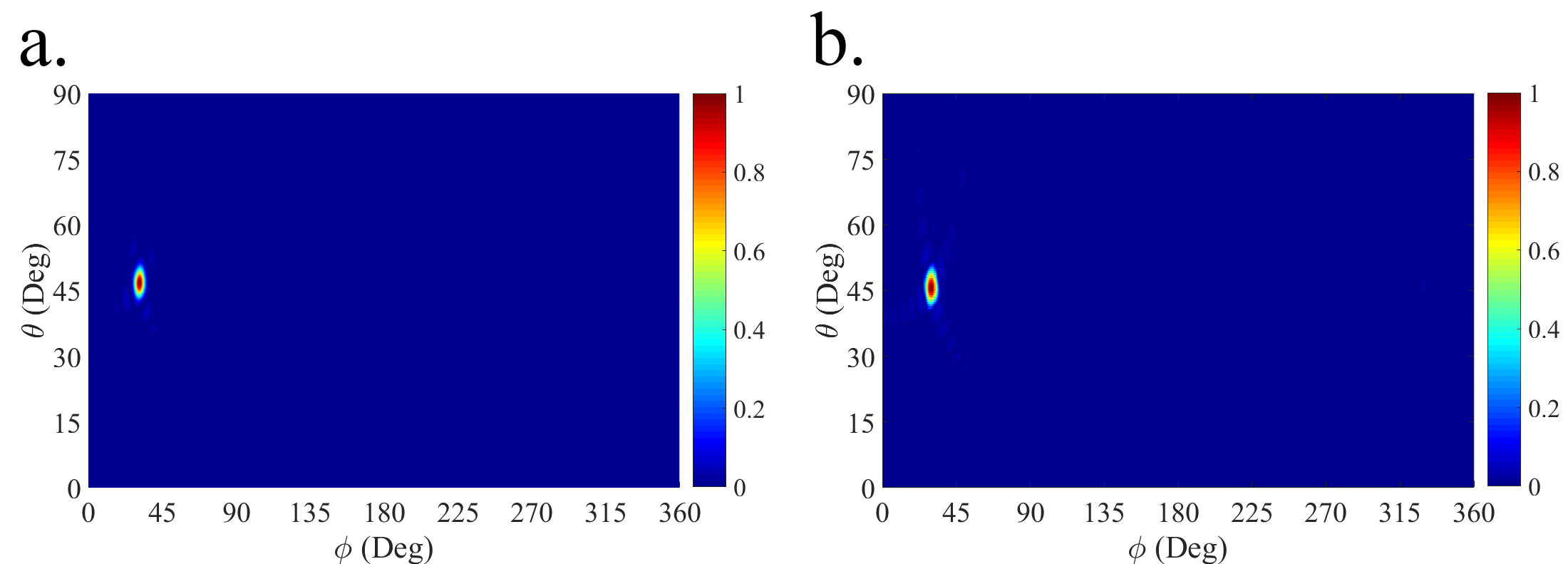}} %,natwidth=426,natheight=538
\vspace{-0.2cm}
\caption{Radiation pattern of the metasurface structure at 2 THz with the main reflected beam pointed at $\{\phi_{r}=30^{o},\theta_{r}=45^{o}\}$  by a) numerical approach and b) analytical approach. The incident wave is normal to the metasurface.
\label{fig:Meta1}}
\vspace{-0.2cm}
\end{figure*} 

\begin{figure*}[!t] 
\centering
\label{fig:Meta22}{\includegraphics[width=1\textwidth]{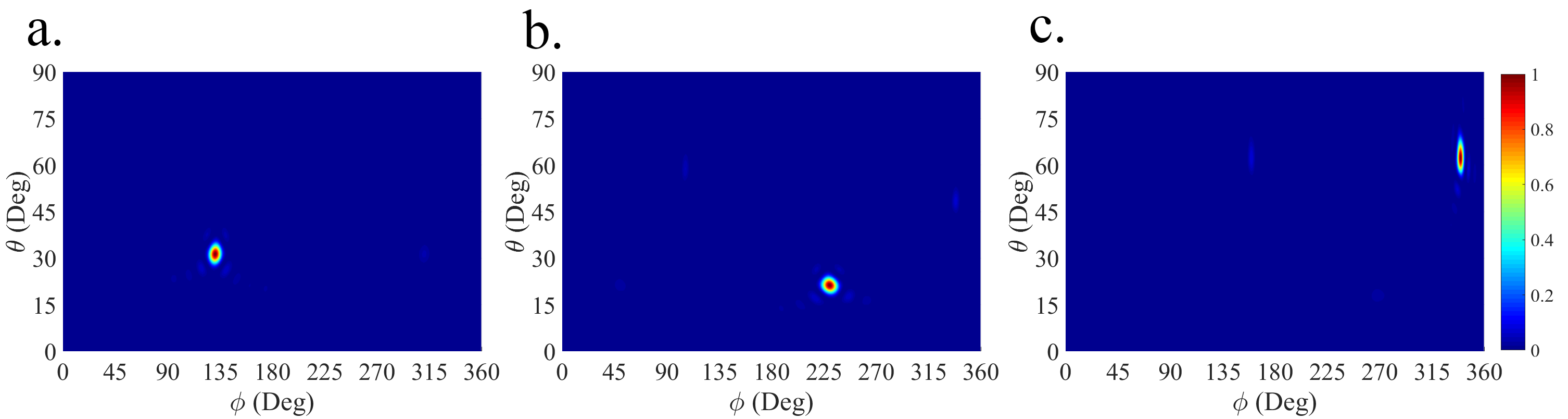}} %,natwidth=426,natheight=538
\vspace{-0.5cm}
\caption{Radiation pattern of the metasurface structure at 2 THz with the main reflected beam pointed at the directions (numerical approach): a) $\{\phi_{r}=130^{o},\theta_{r}=30^{o}\}$, b) $\{\phi_{r}=230^{o},\theta_{r}=20^{o}\}$, and c) $\{\phi_{r}=340^{o},\theta_{r}=60^{o}\}$. The incident wave is normal to the metasurface.
\label{fig:Meta2}}
\vspace{-0.2cm}
\end{figure*} 

\begin{table*}[!t] 
\caption{Design and Performance Results.}
\vspace{-0.2cm}
\label{tab:performance}
\footnotesize
\centering
\begin{tabular}{|c||cc|cc|cc|cc|} 
\hline
Direction & $d_{cx}$ & $d_{cy}$ & $c_{x}$ & $c_{y}$ & $\phi_{3dB}$ & $\theta_{3dB}$ & $Err_{\phi}$ & $Err_{\theta}$  \\ 
$\{\phi_{r}, \theta_{r}\}$ & \multicolumn{2}{c|}{($\upmu$m)} & \multicolumn{2}{c|}{(cells)} & \multicolumn{2}{c|}{(\textsuperscript{o})} & \multicolumn{2}{c|}{(\%)} \\ \hline
\{30\textsuperscript{o}, 45\textsuperscript{o}\} &  61.24  & 106.07 &  3 & 5 & 5 & 5.25  & 2.5  & 3.3 \\
\{130\textsuperscript{o}, 30\textsuperscript{o}\} &  116.68 & 97.91  & 6 & 5 & 8.25 & 4.75 & 0.58 & 4.16  \\ 
\{230\textsuperscript{o}, 20\textsuperscript{o}\} &  170.57 & 143.13 & 8 & 7 & 10.5 & 4 & 0.11 & 6.25 \\
\{340\textsuperscript{o}, 60\textsuperscript{o}\} &  46.08  &  126.6 & 2 & 6 & 4.5 & 8.25 & 0.15 & 3.75 \\
\hline
\end{tabular}
\vspace{-0.3cm}
\end{table*}

In the analytical approach that is used to verify the numerical results, the reflection phase $\Phi(p,q)$ of each unit cell of size $d_{u}$ is assumed to be exactly either 0, $\pi/2$, $\pi$, or $3\pi/2$. Assuming a designed phase distribution assigned to the unit cells, we can express the far-field scattering pattern $F(\theta,\phi)$ as
\begin{equation}
F(\theta,\phi) = f_{E}(\theta,\phi)\times f_{A}(\theta,\phi)
\end{equation}
where $\theta$ and $\phi$ are the elevation and azimuth angles of an arbitrary direction, respectively, and $f_{E}(\theta,\phi)$ and $f_{A}(\theta,\phi)$ are the element factor (pattern function of unit cell) and array factor (pattern function of unit cell arrangement), respectively. Here, the unit cells are assumed to be isotropic, and therefore the scattering pattern depends only on the array factor
\begin{equation}
\begin{array}{l}
F(\theta,\phi) = \sum_{p=1}^{M}\sum_{q=1}^{N}exp\{-j[\Phi(p,q)+ \\
+\, kd_{u}(p-1/2)\sin{\theta}\cos{\phi}\, + \\
+\, kd_{u}(q-1/2)\sin{\theta}\sin{\phi}]\}.
\end{array}
\end{equation}

We first evaluate the metasurface when configured to steer the beam at $\{\phi_{r}=30^{o},\theta_{r}=45^{o}\}$ with $M=N=100$. Following the design flow from Section \ref{sec:flow} and assuming 20-$\upmu$m unit cells and 2-bit coding, we obtain that $d_{cx} = 61.24 \,\upmu$m and $d_{cy} = 106.07 \,\upmu$m, which leads to a cluster of $3\times 5$. The super unit cell thus extends for $12\times 20$ unit cells. Figure \ref{fig:Meta1} shows the far field pattern of the resulting metasurface, which confirms that there is a good agreement between the numerical and analytical solutions. The reason for the small differences between the numerical and theoretical results in sidelobe levels could be the marginal unit cells in the clusters, super unit cells, and the whole structure. We obtain the amplitude and phase of the reflection coefficient of the proposed unit cells while taking into consideration mutual coupling between the same adjacent unit \hl{cells} by assuming a periodic boundary condition. While in the real metasurface structure for beam steering, there are some marginal unit cells which their adjoining unit cells are not similar with them. Consequently, the peridicity condition is broken. In the simulation of full structure, the correct coupling of marginal unit cells is considered while in the theoretical analysis of entire metasurface structure, it is ignored perforce. It is seen that the reflected beam indeed points to the target direction. The steering error, evaluated as the difference between the target and achieved angles, is 2.5\% and 3.3\% in $\phi$ and $\theta$, respectively. The 3-dB width of the beam is approximately 5\textsuperscript{o} in both cases.

To further verify the validity of the approach, we reconfigure the metasurface to operate at three different steering directions. Figure \ref{fig:Meta2} shows how the proposed metasurface design is capable of achieving the desired responses and Table \ref{tab:performance} summarizes the characteristics and performance of the resulting configurations. A wide range of reflected angles is achieved with clusters of 2--8 unit cells, achieving in all cases beam widths below 11\textsuperscript{o} (minimum 4\textsuperscript{o}) with steering error below 7\% (minimum 0.11\%). Note that the error and beam width generally increase when approaching \emph{forbidden areas} in the design space, where the gradient tends to zero or infinity. Also, the reflected beam for the cases \{30\textsuperscript{o}, 45\textsuperscript{o}\} and \{340\textsuperscript{o}, 60\textsuperscript{o}\} especially the latter) tend to be elliptical due to the larger difference between $d_{cx}$ and $d_{cy}$, as hinted in Section \ref{sec:flow}. In addition, to achieve a continuous beam scanning ability of coding metasurface with minimum angle variation, the convolution approach can be leveraged to steer the far-field pattern to a predetermined direction \cite{liu2016convolution}. Regarding the Fourier relation between the field distribution on the coding metasurface and the resultant scattering pattern in the far-field, one can shift the reflected pencil beam by adding the two calculated phase gradient coding \hl{patterns} so that the total phase gradient deflection angle is equal to the desired angle.
%the cluster sizes are $116.68 \times 97.91\,\upmu$m\textsuperscript{2}, $170.57 \times 143.13\,\upmu$m\textsuperscript{2}, and $46.08 \times 126.6 \,\upmu$m\textsuperscript{2}; these dimensions lead to clusters of $6\times 5$, $8\times 7$, and $2\times 6$ unit cells; the beam width is \hl{XXX}; while the steering error is \hl{XXX}. 
%\begin{figure*}[!ht] 
%\centering
%\label{fig:Meta20-230}{\includegraphics[width=1\columnwidth,natwidth=426,natheight=538]{./figures/9.PNG}}
%\vspace{-0.1cm}
%\caption{Metasurface shematic.}
%\vspace{-0.5cm}
%\end{figure*} % You must have at least 2 lines in the paragraph with the drop letter

\section{Programmability and Implementation Issues}
\label{sec:antenna}
The final steps in the design of our beam steering device relate to the elements that control and excite the metasurface. More precisely, we need to conceive a setup that takes the target reflected angle as input and modifies the metasurface accordingly. To this end, in Sec. \ref{sec:controller} we propose a controller that automatically converts the target reflected angle into a bit matrix defining the states of each unit cell. Then, in Sec. \ref{sec:biasing} we discuss the biasing scheme required to address each unit cell with the appropriate voltage (chemical potential). Finally, we review source considerations in Sec. \ref{sec:source}.

\subsection{Controller Design}
\label{sec:controller}
To achieve programmability, it is necessary to attach the metasurface to a digital device capable of translating the beam steering requirements into the global metasurface state. Algorithm \ref{alg1} shows a pseudocode that exemplifies this function. The process starts by calculating the size of the unit cell clusters $c_{x}\times c_{y}$ as a function of the bit number per cell $n$, and the dimension of the unit cell $d_{u}$. Then, the gradient can be built easily by assigning consecutive states to adjacent clusters of unit cells. As already mentioned in Section \ref{sec:flow}, \eqref{eq:dphi3} and \eqref{eq:cluster} can produce negative values, in which case the order of states is reversed. 

%Here, it is worth noting that the results may be negative, which basically implies that the gradient needs to be reversed at the coding stage. Also, regardless of the number of bits, $d_{cx}$ and $d_{cy}$ may become arbitrarily large as the target direction approaches $\phi_{r} = 0, \theta_{r} = 0$. This is, indeed, consistent with reality as such direction implies specular reflection, which can only be realized with a homogeneous surface, i.e. zero gradient. 
Algorithm \ref{alg1} assumes that all unit cells are addressed by a centralized device, probably an FPGA. However, since the metasurface implements a discretized gradient, it would be relatively straightforward to come up with an algorithm that can calculate the required state in a distributed way, only relying on the state of the immediate neighbour. Such simplified scheme would be suitable for the rising Software-Defined Metamaterial (SDM) paradigm \cite{Liaskos2015, AbadalACCESS}, which aims to provide programmable metamaterials that can be reconfigured via an integrated network of controllers that drive unit cells individually. In that case, an external entity called gateway would receive the command of changing the direction of the beam. The gateway would compute $c_{x}$ and $c_{y}$, then rely them to the first controller together with $n$. The first controller would be initialized and pass its state along with $c_{x}$, $c_{y}$, and $n$ to their neighbours, which would repeat the process until the whole metasurface is programmed.

\subsection{Actuator Design}
\label{sec:biasing}
\begin{figure}[!t] 
\centering
{\includegraphics[width=0.7\columnwidth]{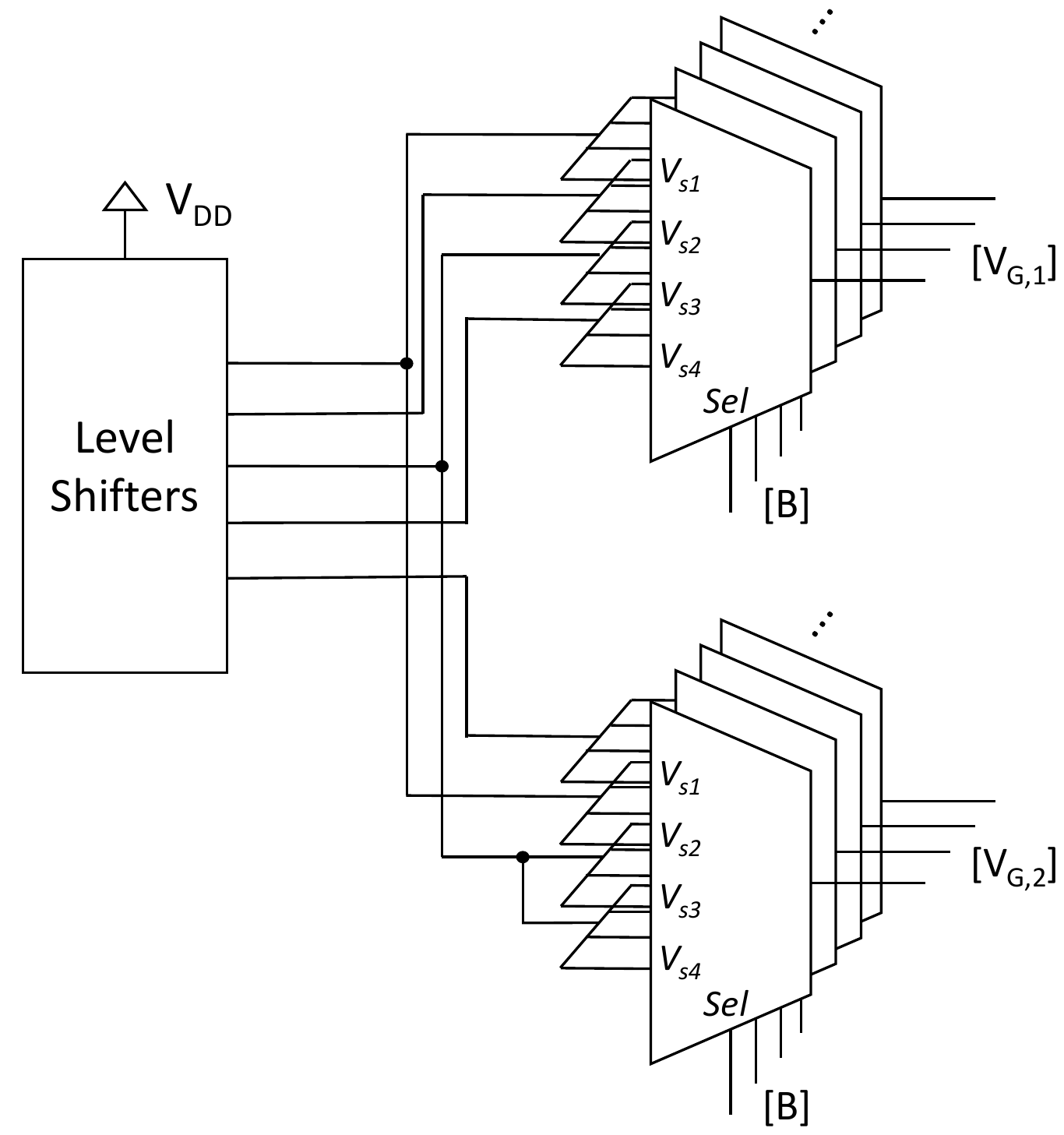}}
\vspace{-0.3cm}
\caption{Sample implementation of the actuator for the metasurface based on the $2G$ unit cell and 2-bit coding.\label{fig:actuator}}
\vspace{-0.3cm}
\end{figure}

%Once the state matrix is formed, an actuator uses the digital output coming from the controller to deliver the appropriate voltage $v_{g}(i) \in [V_{G}]$ to each metasurface element. 
%In other words, the actuator translates the state matrix onto the voltage levels that will achieve the required chemical potentials in each graphene patch of the metasurface. 
The actuator is a circuit that translates the state matrix $[B]$ provided by the controller into the matrix of appropriate voltages $[V_{G}]$ that, in turn, leads to the required chemical potentials in each graphene patch of the metasurface. As shown in Fig. \ref{fig:actuator}, a set of voltage level shifters and a matrix of multiplexers would be enough for this purpose. Note that several independent sets of multiplexers (two in our case) may be required to drive the graphene patches of individual unit cells. It is also worth noting that only five distinct voltages are needed in our case, because several states share the same target chemical potentials according to the calculations made in Section \ref{sec:states}.

\begin{algorithm}[!t]
\caption{Algorithm for clustered gradient formation.}
\label{alg1}
\begin{algorithmic}
\small
\STATE Inputs: du, phiR, thetaR, n, f
\STATE
\STATE /* CALCULATION OF THE CLUSTER SIZES */
\STATE lambda = 3e8/f;
\STATE dcx = lambda/(2\^{}n*cos(phiR)*sin(thetaR));
\STATE dcy = lambda/(2\^{}n*sin(phiR)*sin(thetaR));
\STATE cx = round(dcx/du); // MAY BE NEGATIVE
\STATE cy = round(dcy/du); // MAY BE NEGATIVE
\STATE
\STATE /* CALCULATION OF THE STATE MATRIX */
%\STATE B(0,0) = 0; 
\STATE for(i=1; i$<$M; i++) \{
\STATE $\,\,\,\,$for(j=1; j$<$N; j++) \{
\STATE $\,\,\,\,\,\,\,\,$B(i,j) = (round(i/cx) + round(j/cy)) mod(2\^{}n);
\STATE $\,\,\,\,$\}
\STATE \}
\end{algorithmic}
\end{algorithm}

The actual voltages required at the output of the level shifters mainly depend on the graphene biasing structure and the required chemical potential \cite{Huang2012ARRAY, Gomez2015}. The configuration assumed in this paper is similar to that used in \cite{Huang2012ARRAY}, which couples graphene capacitively with a back gate through a thin Al$_{2}$O$_{3}$ layer. Essentially, this scheme shifts the operation of graphene between the Dirac point, where the chemical potential is minimum, and a point where the potential reaches the maximum desired point. The resulting chemical potential $\mu_{c}$ relates to the change of voltage $\Delta v_{g}$ as 
\begin{equation}
\label{eq:SLG}
\Delta v_{g} = \frac{e \mu_{c}^2 t}{\pi \hbar^2 v_{F}^2 \varepsilon_{0} \varepsilon_{r}},
\end{equation}
where $e$ is the elementary charge, $\hbar$ is the reduced Planck constant, $v_{F} \approx 10^{6} $ m/s is the Fermi velocity, $\varepsilon_{0}$ is the vacuum permittivity, whereas $\varepsilon_{d}$ and $t$ are the permittivity and thickness of the material below graphene \cite{Yu2009Chemical}. 

\begin{figure*}[!ht] 
\centering
\subfigure[Voltage as a function of the dielectric thickness.\label{fig:voltT}]{\includegraphics[width=1\columnwidth]{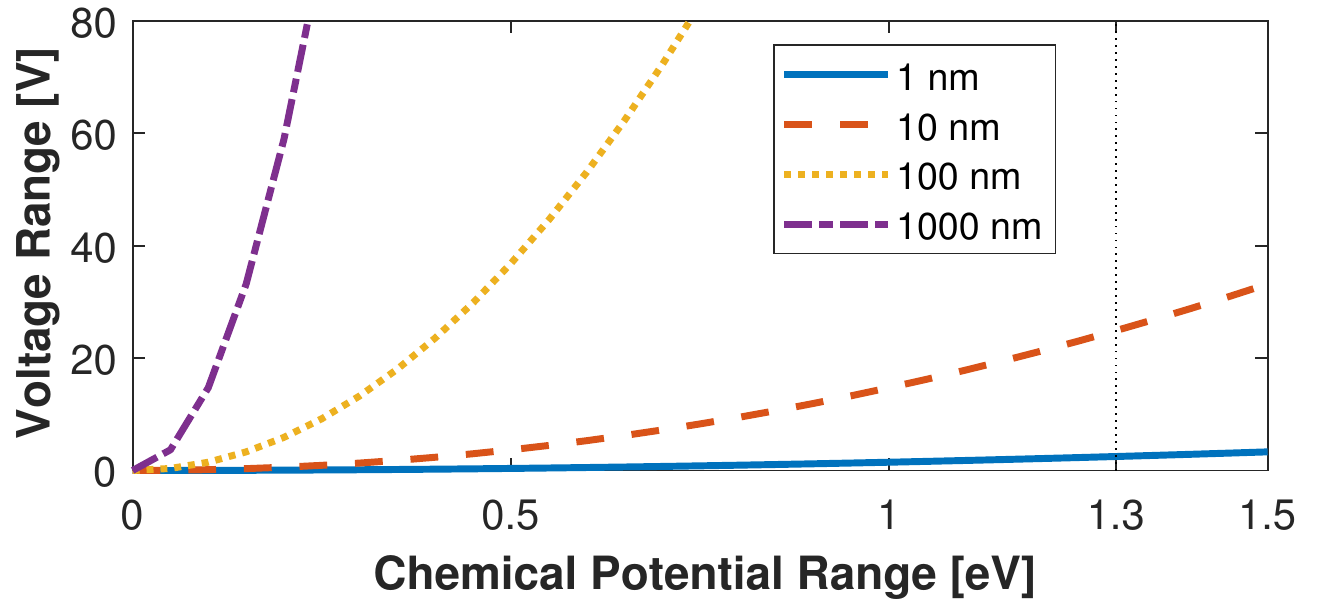}} 
\subfigure[Voltage as a function of the dielectric constant.\label{fig:voltE}]{\includegraphics[width=1\columnwidth]{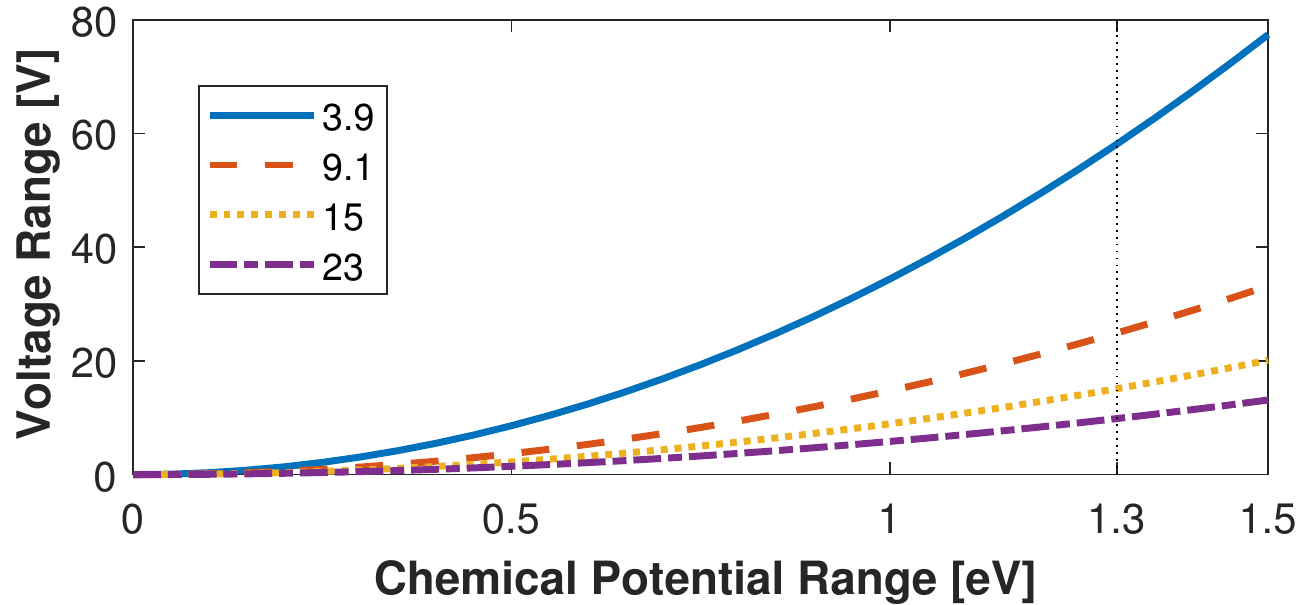}} 
\vspace{-0.1cm}
\caption{Voltage range required at the level shifting stage to achieve a given target chemical potential range.\label{fig:voltage}}
\vspace{-0.3cm}
\end{figure*}

Figure \ref{fig:voltage} illustrates the voltage ranges required to achieve a certain target chemical potential range. As directly implied by \eqref{eq:SLG}, the voltage requirements increase quadratically with the target chemical potential range. To limit the requirements, one can either minimize the space between the gate and the graphene layer or use materials with high dielectric constant. However, the former is determined by technological constraints, and the latter needs to take into consideration the cost and other characteristics of the material.

%The relation $v_{g}$--$E_{F}$ becomes linear for bilayer graphene and remains unclear for trilayer graphene and above. In any case, the scheme here presented is applicable to any graphene structure capable of being reconfigured through electrostatic biasing.
% FIX: Sergi - other types of biasing? Top-direct gating?

The chemical potential range required by our metasurface can be obtained easily once the unit cell states are defined. In the present design, $\Delta\mu_{c} = 1.3$ eV. Assuming a Al$_{2}$O$_{3}$ layer ($\varepsilon_{r} = 9.1$) with thickness $t = 10$ nm, achievable with current technologies \cite{Huang2012ARRAY}, the resulting voltage range is 24.9 V. 

\subsection{Towards an Experimental Setup}
\label{sec:source}
\begin{figure}[!ht] 
\centering
{\includegraphics[width=0.8\columnwidth]{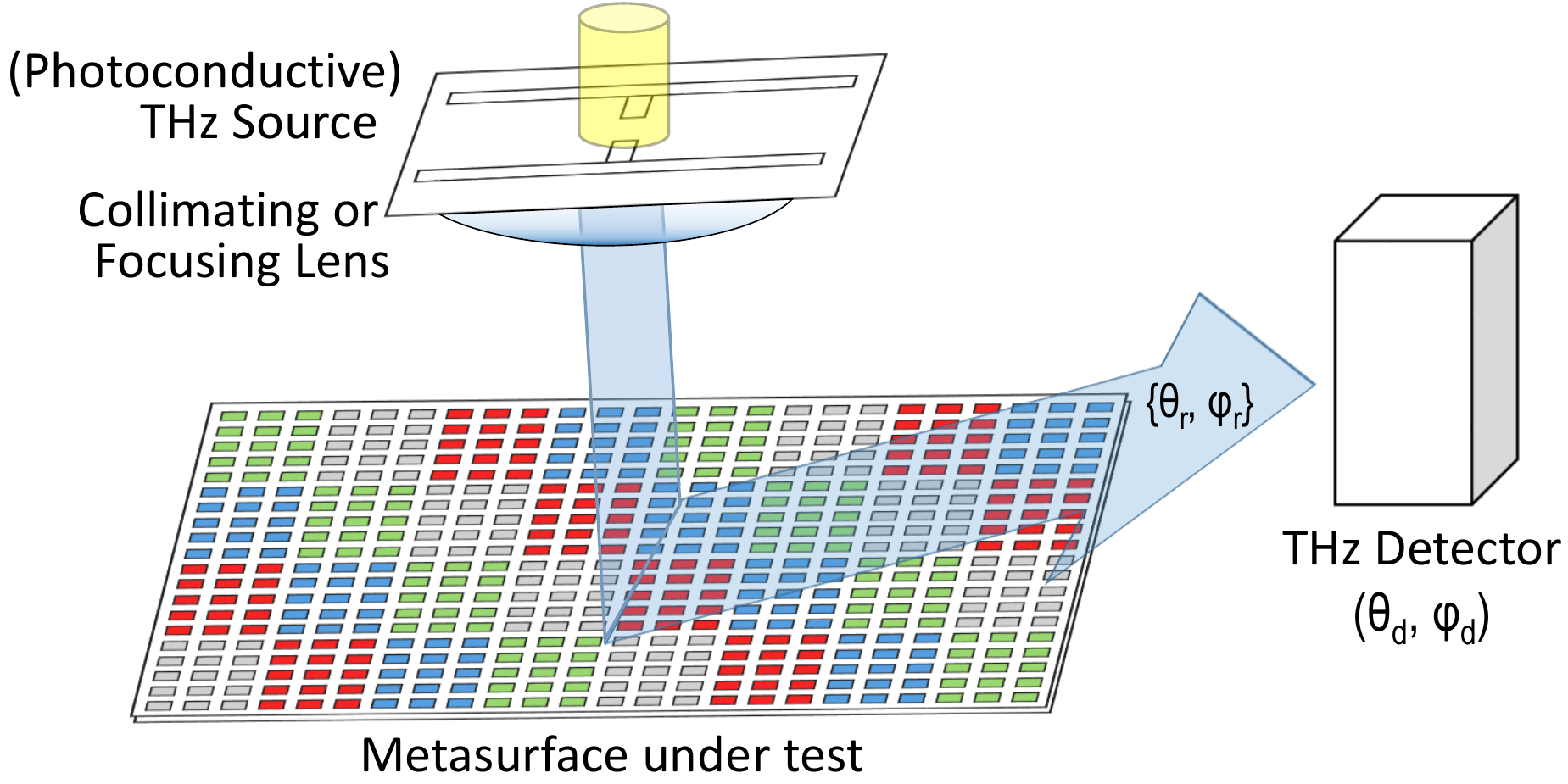}}
\vspace{-0.3cm}
\caption{\hl{Sketch of a potential measurement setup for the proposed device.}\label{fig:setup}}
\vspace{-0.3cm}
\end{figure}

Figure \ref{fig:setup} illustrates a possible measurement setup for the experimental \hl{validation} of the metasurface. \hl{The testbed would mainly consist of a fiber-coupled time domain spectroscopy (THz-TDS) system with a fixed source (generally harder to calibrate) and a movable receiver placed on a rotatory platform. The source is based on a photoconductive antenna coupled to a focusing or collimating lens that minimizes spreading losses. Additional optics such as parabolic reflectors can be incorporated to meet the receiver sensitivity as well as the source--metasurface--receiver distance requirements. It is worth noting that such a scheme has been used successfully in other works proving anomalous reflection in the THz band} \cite{Liu2016a, Liang2015a}\hl{. A very similar system has been built in} \cite{Kokkoniemi2016} \hl{for the measurement of the reflection coefficient of surfaces in the terahertz band entirely with commercial solutions.}

\section{Discussion}
\label{sec:disc}
In this section, we qualitatively discuss several cross-cutting issues related to the design of the metasurface.

\begin{figure*}[!ht] 
\centering
\subfigure[Achieved $\theta_{r}$ for different values $d_{u}$.\label{fig:rangeDX}]{\includegraphics[width=0.68\columnwidth]{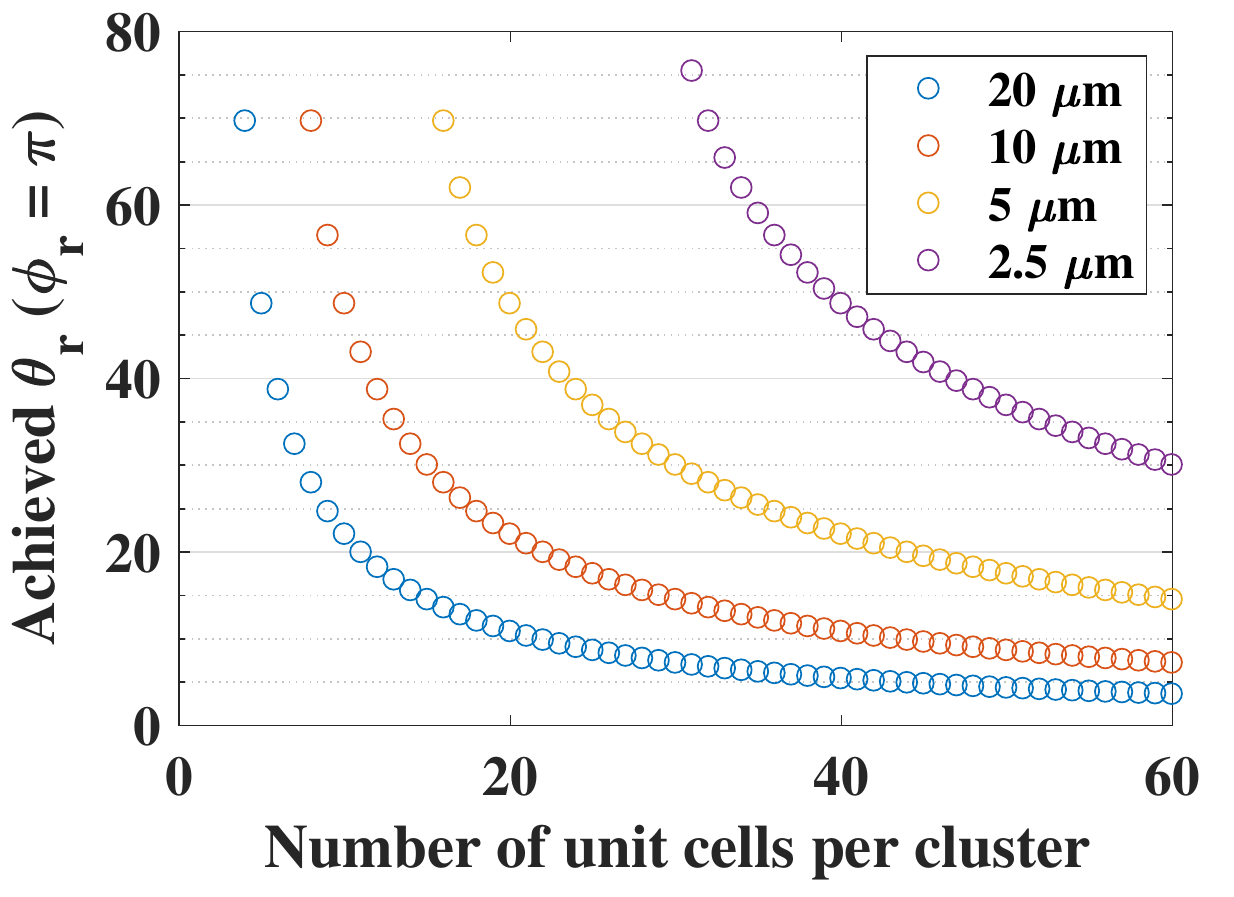}}
\subfigure[Far field patterns for different number of bits (top) and metasurface sizes (bottom).\label{fig:resDX}
%\textcolor{red}{[Please add unit for x,y-labels. If it is possible, please make it with better quality]}
\label{fig:sizeDX}]{\includegraphics[width=1.3\columnwidth]{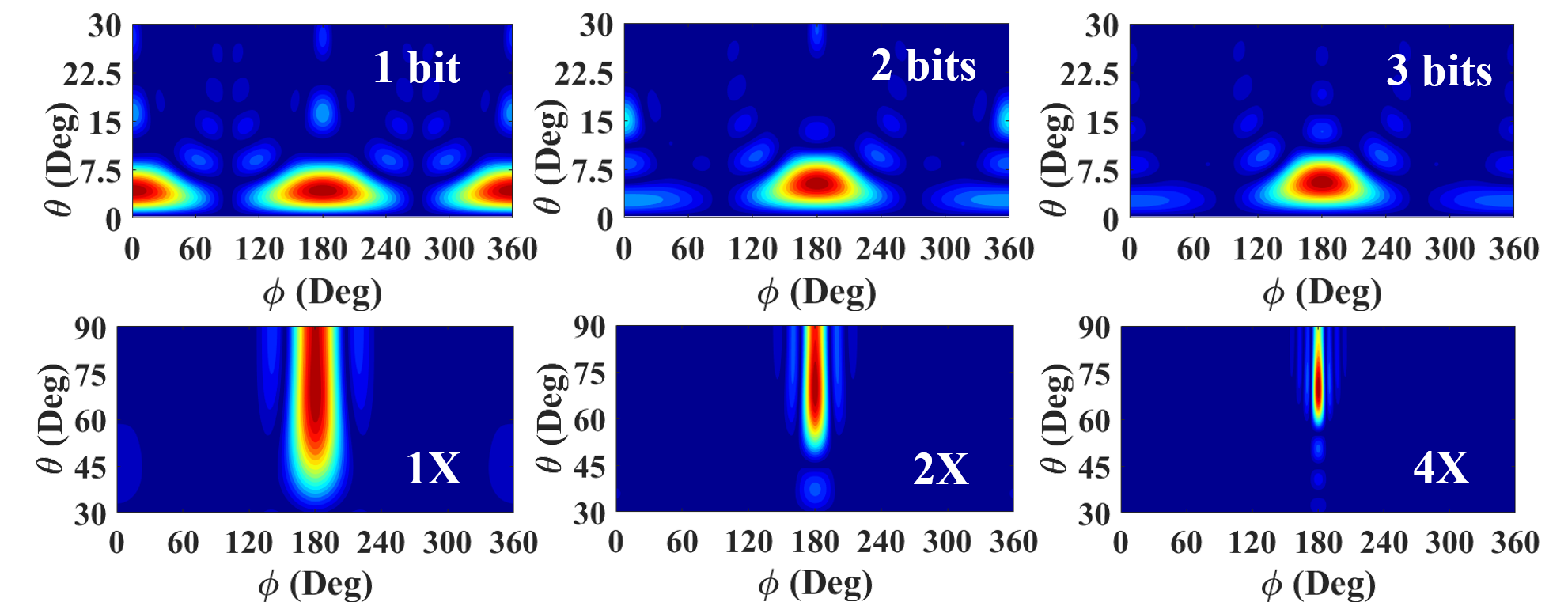}}
%\subfigure[Far field patterns for different metasurface sizes.\label{fig:sizeDX}]{\includegraphics[width=1\columnwidth]{./figures/size_dx.png}}
\vspace{-0.1cm}
\caption{Performance scalability analysis of the metasurface from the unit cell, coding, and complete device perspectives.}
\vspace{-0.3cm}
\end{figure*}

\vspace{0.1cm}
\noindent
\textbf{Scalability analysis:} To accommodate the proposed design flow to different beam steering specifications, it is crucial to understand which are the key design parameters and what performance metrics do they affect. Here, we highlight some:

%More bits - narrower beam, better spatial resolution. Quantification of the resolution, see if it enough.
\begin{itemize}
\item The size of the unit cell presents an interesting tradeoff. While it may be difficult to achieve a wide phase range if the unit cell is too small with respect to the wavelength (see Figs. \ref{fig:results1G} and \ref{fig:results1Gp}), reducing its dimensions leads to a raise in the maximum achievable phase gradient. This is useful to achieve better control at the end-fire directions ($\theta_{r}\to 90$\textsuperscript{o}) of the metasurface, as exemplified by Fig. \ref{fig:rangeDX}. For instance, we can achieve beam steering at $\theta_{r} > 70$\textsuperscript{o} only for $d_{u} < 5 \upmu$m. For angles even closer to $\theta_{r}=90$\textsuperscript{o}, a design converting the incident wave into a surface wave may be required \cite{Tcvetkova2017}. In any case, note that such fine-grained control at THz frequencies can only be achieved with graphene, thanks to its support of plasmonic slow-wave propagation in this frequency band. 

%- smaller unit cells make it harder to achieve the phase range. However, they allow to apply larger gradients (phase change is fixed, but dx diminishes). This means better control for directions close to infinite gradient.
%%%% range2_dx.m
\item Increasing the number of bits provides better control on the phase as it allows to draw the phase gradient more accurately, with more clusters and less unit cells per cluster. This is of special importance in directions close to the boresight ($\theta_{r}\to 0$\textsuperscript{o}), where a subtle gradient is required. Fig. \ref{fig:resDX} exemplifies this for a design targeting $\theta_{r} = 5.37$\textsuperscript{o} with fixed size, but increasing number of bits. The 3-bit instance greatly reduces the side lobes and has its maximum at $\theta_{r} = 5.26$\textsuperscript{o}, whereas the beam moves away from the desired direction for the 2-bit and 1-bit cases (4.92\textsuperscript{o} and 4.22\textsuperscript{o}, respectively). Note, however, that the gain in accuracy comes at the cost of a substantially higher complexity at the controller and the actuator. 
%- higher number of bits provides better control on the phase. In the end, that may increase the accuracy of the steering (gradient is better defined). It also may increase control for directions close to zero gradient, in which we need a subtle gradient.
%%%% resolution_dx.m
\item In arrays, adding more antennas allows to reduce the beam width. The same principle should apply in our design, as exemplified in Section \ref{sec:evaluation} with the array factor formulation. We verified such hypothesis by fixing the gradient and doubling the number of supercells once and twice (from $M=N=40$ to $M=N=160$). The resulting far field patterns, shown in Fig. \ref{fig:sizeDX}, clearly demonstrate that the beam is sharpened without significantly changing the direction of maximum energy. In fact, the beam width is reduced by a factor proportional to the increase in metasurface size, i.e. from 50\textsuperscript{o} and 5\textsuperscript{o} to 12\textsuperscript{o} and 1.3\textsuperscript{o} in the $\theta$ and $\phi$ angles, respectively.
%Clearly, the ene increasing the size of the metasurface helps improving the  \hl{xxx}. >> no change in maximum direction.
%%%% size_dx.m
\end{itemize}

\vspace{0.1cm}
\noindent
\textbf{Co-design opportunities:} Understanding the design flow from the point of view of the unit cell, the metasurface, or the device as a whole helps to identify possible co-design opportunities. For instance, the unit cell configuration and the coding determine the complexity in terms of number of required voltage levels, as well as the quantity and size of the multiplexers. For an $n$-bit coding with $g$ independently biased graphene patches, we may require up to $g\cdot 2^{n}$ levels and $g$ multiplexers with $2^{n}$ inputs. However, advanced design exploration techniques may allow to find design points that reduce the voltage range, the number of required levels, and the multiplexer inputs while gracefully degrading the performance of the system. For instance, in our 2-bit implementation with the $2G$ unit cell, we could take $\mu_{c,1} = \mu_{c,2} = \{0.1, 0.6\}$ and still achieve reasonable performance, but with a $4\times$ reduction of voltage range and number of levels and greatly simplfying the multiplexing circuits.
%After understood from the point of view of the unit cell, metasurface, and controller, we 
%The unit cell configuration and the coding determine the complexity (number of voltages, number of contacts...). One can simplify it by reducing the number of bits to the minimum required by the application, looking for design points that reduce the voltage range while gracefully degrading the performance of the system, or also looking for a reuse of the voltage levels in the different points  (example, in the $2G$ unit cell, taking points in a square so that each bit addresses only one layer).
%Choose the points to minimize the chemical potential range. Maybe increasing relaxation time?

\vspace{0.1cm}
\noindent
\textbf{Adaptive clusterization:} The strength of the proposed design flow is its simplicity. By fixing the cluster size first and then statically building the super cells, the state matrix can be calculated easily. However, the rounding operation used in the number of unit cells per cluster (Equation \eqref{eq:cluster}) introduces an error, specially for large unit cells, that is later amplified by the static building of super cells. Both issues can be alleviated by simply inverting the design flow, i.e. obtaining the size of the super cell first, and then breaking it down into unequal clusters. For instance, a super cell composed by 18 unit cells can be coded with 4, 5, 4, and 5 unit cells per state; otherwise, the super cell would be statically coded to either 4 or 5 unit cells per state, leading to a significant error. In a similar approach, the coding algorithm could dynamically adapt the number of states, using fewer bits in those directions that require a very large gradient.
%The size of the unit cell, as a discretization of the space, also yields interesting tradeoffs. Smaller unit cells with respect to the wavelength reduce the phase range, and thus reduce their practicality for beam manipulation purposes. However, coarse unit cells lead to an increasing error since we need 	A way to combat this error is through the non-uniform assignment of unit cells in the super unit cell. Instead of deriving $s_{x}$ and $s_{y}$ as multiples of $c_{x}$ and $c_{y}$, one can derive the super unit cell from the target reflected direction and then break it down to the clusters.

\section{Conclusion}
\label{sec:conclusion}
This paper has presented the complete design, from the unit cell up to the programming algorithm, of a reconfigurable digital metamaterial for beam steering in the terahertz band. The tunability of graphene is exploited at the unit cell level to provide a phase range close to $2\pi$, whereas the generalized Snell's law of reflection has been used to derive the phase gradients required to target the beam to the desired direction. The results confirm the validity of the approach, which for normal incidence achieves a very broad reflection range with angle-dependent beam widths and steering errors. Considering normal incidence, the analytical formulation also models forbidden (and unreasonable) reflection directions effectively as infinite gradients. Finally, the scalability analysis confirms that the beam width depends on the size of the metasurface, the reflection range depends on the size of the unit cells, and the steering error and side lobe levels depend on the number of phases that the graphene-based unit cells can implement. Future works could leverage the comprehensive methodology developed herein to optimize the unit cell design and phase gradient formation to reduce the overhead of the solution and further improve the beam steering performance.

% or


%%%\appendices
%%%\section{Proof of the First Zonklar Equation}
%%%Appendix one text goes here.

%%%\section{}
%%%Appendix two text goes here.

\section*{Acknowledgment}
This work has been partially funded by Iran's National Elites Foundation (INEF), the Spanish Ministry of \emph{Econom\'ia y Competitividad} under grant PCIN-2015-012, and by ICREA under the ICREA Academia programme. Also, the authors would like to thank Christoph S{\"u}{\ss}meier and the anonymous reviewers for their invaluable feedback.%, and the European Union via the Horizon 2020: Future Emerging Topics call (FET Open), grant EU736876, project VISORSURF (http://www.visorsurf.eu).

%%%\ifCLASSOPTIONcaptionsoff
%%%  \newpage
%%%\fi

\section*{Appendix}
To extract $\phi_{r}$ from \eqref{eq:dphi}, we divide both expressions and apply basic trigometry to obtain
\begin{equation}
\tan{\phi_{r}} = \frac{k_{i} \sin{\theta_{i}}\sin{\phi_{i}} + \frac{d\Phi}{dy}}{k_{i} \sin{\theta_{i}}\cos{\phi_{i}} + \frac{d\Phi}{dx}}  
\end{equation} 
which yields 
\begin{equation}
\phi_{r} = \arctan{\frac{k_{i} \sin{\theta_{i}}\sin{\phi_{i}} + \frac{d\Phi}{dy}}{k_{i} \sin{\theta_{i}}\cos{\phi_{i}} + \frac{d\Phi}{dx}}} %= \arctan{\frac{k_{iy}+\nabla\Phi_{y}}{k_{ix}+\nabla\Phi_{x}}}
\end{equation} 

To extract $\theta_{r}$ from \eqref{eq:dphi}, we square and sum both expressions:
\begin{equation}
\begin{array}{l}
k_{r}^{2} \sin{\theta_{r}}^{2}\cos{\phi_{r}}^{2} + k_{r}^{2} \sin{\theta_{r}}^{2}\sin{\phi_{r}}^{2} = \\
= (k_{i} \sin{\theta_{i}}\cos{\phi_{i}} + \frac{d\Phi}{dx})^{2} + (k_{i} \sin{\theta_{i}}\sin{\phi_{i}} + \frac{d\Phi}{dy})^{2}  
\end{array}
\end{equation} 
which, after applying basic trigonometry, becomes
\begin{equation}
\begin{array}{l}
k_{r}^{2} \sin{\theta_{r}}^{2} = \\
= (k_{i} \sin{\theta_{i}}\cos{\phi_{i}} + \frac{d\Phi}{dx})^{2} + (k_{i} \sin{\theta_{i}}\sin{\phi_{i}} + \frac{d\Phi}{dy})^{2}
\end{array}
\end{equation} 

Isolating, we obtain
\begin{equation}
\begin{array}{l}
k_{r} = \arcsin{ \frac{\sqrt{(k_{i} \sin{\theta_{i}}\cos{\phi_{i}} + \frac{d\Phi}{dx})^{2} + (k_{i} \sin{\theta_{i}}\sin{\phi_{i}} + \frac{d\Phi}{dy})^{2} }}{k_{r}} } %= \\
%= \arcsin{\frac{\sqrt{(k_{ix}+\nabla\Phi_{x})^{2} + (k_{iy}+\nabla\Phi_{y})^{2}}}{k_{r}}} 
\end{array}
\end{equation}


\end{document}